\documentclass[aps,prb,twocolumn,groupedaddress,showpacs]{revtex4-1}
\usepackage{mathbbol,amsmath,amsfonts,amssymb,bm,color,dsfont,bbold}
\usepackage{graphicx,psfrag,array}

\providecommand{\ignore}[1]{}
\providecommand{\aucmnt}[1]{#1}
\def\be{\begin{equation}}
\def\ee{\end{equation}}
\renewcommand{\aucmnt}[1]{}

\newcommand{\ket}[1]{| #1 \rangle}

\newcommand{\Comment}[1]{}

\newcommand{\Eq}[1]{Eq.~(\ref{#1})}

\begin{document}

%\title{Rigorous relation between Pairing and Quantum Hall Physics}
\title{Algebraic approach to the study of zero modes of Haldane pseudopotentials}
\author{Li Chen}
\author{Alexander Seidel}
\affiliation{Department of Physics, Washington University, St.
Louis, MO 63130, USA}

\date{\today}
\begin{abstract}
We consider lattice Hamiltonians that arise from putting Haldane pseudo-potentials
into a second quantized or ``guiding-center-only'' form.
These are fascinating examples for frustration free lattice Hamiltonians.
This is so since even though their highest density zero energy ground states, the Laughlin states,
are known to have matrix-product structure (with unbounded bond dimension), the frustration free character of these lattice Hamiltonians seems obscure, {\em unless} one goes back to the original first quantized picture of analytic lowest Landau level wave functions. This step involves putting back additional degrees of freedom associated with dynamical momenta, and one wonders whether the addition of these degrees of freedom is truly necessary to recognize the frustration free character
of the underlying lattice Hamiltonian. Fundamentally, these degrees of freedom have nothing to do with spectrum of a
``guiding-center-only'' Hamiltonian. Moreover, such constructions are unfamiliar and not available in the study of simpler (finite range) frustration free lattice Hamiltonians with matrix product ground states (of finite bond dimension). That the zero mode properties of ``lattice versions'' of pseudo-potentials can be understood from a polynomial-free, intrinsically lattice point of view is also suggested by the fact that these pseudo-potentials are constructed from an algebra of reasonably simply looking operators. Here we show that zero mode properties, and hence the frustration free character, of these lattice Hamitlonians can be understood as a consequence of algebraic structures that these operators are part of. We believe that our results will deepen insights into parent Hamiltonians of matrix product states with infinite bond dimensions, as could be of use, especially, in the study of fractional Chern insulators.
\end{abstract}
\pacs{73.43.Cd}
\maketitle

\section{Introduction}

Exactly solvable models of quantum mechanical systems serve to corroborate many
of the most fundamental paradigms for the behavior of quantum matter.
While more often than not, one is interested in the behavior of systems that are
far from solvable, powerful effective theories often flow from a deep understanding of  few
isolated special Hamiltonians whose key properties are known exactly.
Similarly, while no exact solutions can be given for the electronic wave functions of any atom
save the simplest, hydrogen, a great wealth of atomic physics, quantum chemistry, and solid state
physics is fundamentally built upon the latter. In the most fortunate cases, the solutions
of such special Hamiltonians can be obtained in a variety of different approaches, each
revealing intricate underlying mathematical structures that may be useful in many contexts
beyond the scope of the original problem.
Already in his original work,\cite{schrodinger} Schr\"odinger gave both the analytical
wave function solution to the harmonic oscillator, rooted in the wealth of knowledge on differential
equations and special functions inherited from the 19th century, as well as an algebraic
construction. It was the latter that has deeply influenced the development of many-body physics
and quantum field theory. A little later, Pauli introduced a more algebraic approach for the hydrogen atom,
emphasizing the role of the symmetry that is associated to the conservation of the Lenz vector.\cite{alg_hydrogen}
Since the very early days, it has been a characteristic of quantum theory that we may often choose between a language of analytic wave functions satisfying differential wave equations and, more generally, an algebraic description, where the fundamental object is the C$^\ast$-algebra of observables. This dichotomy is apparent already in the different pictures of quantum physics associated to the names of Schr\"odinger and Heisenberg.

A particular niche of quantum many-body physics is defined by the description of fractional
quantum Hall states and their rich phenomenology. Special Hamiltonians in the sense described
above have played an important role in this field since Haldane pointed out\cite{haldane_hierarchy}
that Laughlin states are the exact ground states of certain pseudo-potentials, and shortly thereafter, the potential relevant to the  $\nu=1/3$ state was characterized as a Landau level projected ultra-short-ranged interaction by Trugman and Kivelson.\cite{TK} Unlike in other related fields, however, the main focus has been on the construction of first quantized wave functions as pioneered by Laughlin, that are written down quite independent of any Hamiltonian principles and are either required to satisfy certain analytic ``clustering properties'',\cite{Laughlin, MR, RR, BH1, BH2, BH3, WW1, WW2} and/or are obtained from a given conformal edge theory.\cite{MR}
In the most fortunate cases, including Laughlin,\cite{Laughlin} Moore-Read,\cite{MR} Read-Rezayi,\cite{RR} and the Gaffnian state,\cite{gaffnian} the aforementioned analytic properties also lend themselves to the construction of suitable parent Hamiltonians. E.g., in the aforementioned case of the $\nu=1/3$ Laughlin state, it is the characteristic property of the wave function to vanish as the third power of inter-particle distance, whenever two particles are approaching one-another, that allows for the construction of a local parent Hamiltonian. Since their construction is based on analytic many-particle wave functions, these Hamiltonians are usually defined in a first quantized Language.
For example, the Hamiltonian stabilizing the $\nu=\frac{1}{M}$ Laughlin state is given by\cite{haldane_hierarchy}
\be
  H_{\frac 1 M}=\sum_{\substack{0\leq m<M\\ (-1)^m=(-1)^M}}\sum_{i<j} P^m_{ij}\label{1stquant}
\ee
where $i$ and $j$ are {\em particle} indices, and $P^m_{ij}$ projects the pair of particles with
indices $i$ and $j$ onto states with relative angular momentum $m$.
Generalizations of this construction exist, e.g., for the Moore-Read state,\cite{greiter}
the Read-Rezayi series,\cite{RR} and for the Gaffnian.\cite{gaffnian}
In all these cases, one obtains positive Hamiltonians, whose zero energy
ground states (zero modes) must satisfy certain analytic clustering conditions,
and are given by the special wave functions defining ``incompressible'' ground states
for the respective quantum Hall phase, as well as quasi-hole type excitations. The counting statistics
of the latter are fundamentally related to conformal edge theories.\cite{RRcount, milovanovic, ardonne}

In this paper, we discuss and further develop an alternative -- algebraic -- route to the construction of zero modes of the Hamiltonian \Eq{1stquant}, which does not make direct contact with the analytic clustering properties of first quantized wave functions. We believe that this Hamiltonian is of such fundamental interest that the exploration of its inner workings through a different framework will shine new light on the deep mathematical structures underlying fractional quantum Hall states, and may ultimately lend itself to the construction of new parent Hamiltonians. To make the problem concrete, we note that Landau level projection was left implicit in \Eq{1stquant}, but is routinely enforced.
The effect of Landau level projection is to fix the degrees of freedom of the system associated with dynamical momenta, which determine the structure of a given Landau level. This leaves as the effective degrees of freedom the guiding center coordinates, and leads to the usual representation of the Hilbert space as a one-dimensional (1D) ``lattice''. The orbitals associated to this lattice are Landau level states with guiding center coordinates characterized by a single integer, angular momentum-like quantum number. It makes sense to write out the Hamiltonian \eqref{1stquant} in second quantized form,
making explicit the dynamics in this guiding center orbital occupation number basis \cite{CHRB, ortiz}:
\begin{equation}
\begin{split}
    H_{\frac 1 M}=\sum_{\substack{0\leq m<M\\ (-1)^m=(-1)^M}} \sum_R {Q^m_R}^\dagger Q^m_R\,,\\
    \text{where}\qquad Q^m_R=\sum_x \eta_{R,x}^m c_{R-x} c_{R+x} \label{2ndquant}\,.
 \end{split}
\end{equation}
Here, the sum over $R$ is over both integer and half-odd integer values of the ``center of mass''
of a pair of particles destroyed by $Q^m_R$. The sum over $x$ is over integer (half-odd-integer) when $R$ is integer (half-odd-integer). The form factors $\eta_{R,x}^m$ depend on the geometry,
which shall be the disk, sphere, or cylinder geometry for the purpose of this paper.
Traditionally, the form \eqref{2ndquant} for the Haldane pseudo-potentials (and its generalizations for $n$-body interactions) has been given preference mostly for numerical work, as it makes Landau level projection and thus the dimensional reduction of the Hilbert space explicit. Today, however, there is additional motivation to be interested in Hamiltonians of the form given by \Eq{2ndquant}.
The idea of using the second quantized form of quantum Hall type Hamiltonians to generate
frustration free lattice models for exotic electronic states in solids has been advocated by Lee and Leinaas,\cite{LL} and also in Ref. \onlinecite{seidel05}. Here, the orbital basis acted upon by the operators $Q^m_R$ are Wannier states. Additional indices can be added to make such models describe systems in more than one dimension. However, as pointed to by Qi,\cite{qi}
a natural mapping exists between Wannier states of two-dimensional Chern band and
Landau level orbitals in the cylinder geometry. Such Chern bands, in particular if they are
flat\cite{Tang11,Sun11,Neu11,Sheng11,Wang11,Regnault11,Ran11,Bernevig12,WangPRL12,WangPRB12,Wu12,Laeuchli12,Bergholtz12,Sarma12,Liu13,chen13, scaffidi} (though strictly, this requires\cite{chen13} non-local hopping terms),
together with appropriate interactions may harbor the sought-after fractional Chern insulator.
Common to all these applications in solids is the fact that the first quantized versions of the
respective Hamiltonians, e.g., \Eq{1stquant} and the analytic forms of traditional quantum Hall ground states, are essentially meaningless; only second quantized forms, such as, \Eq{2ndquant} and a purely ``guiding center'' presentation of the wave function have natural meaning.
For these reasons, there is much renewed interest in the ``lattice'' variant of quantum Hall-type
Hamiltonians, especially\cite{CHLee} the manifestly translationally invariant type associated to the cylinder geometry.

Moreover, it has recently been argued by Haldane\cite{Haldane11} that the essence of quantum Hall states such as Laughlin states lies in their guiding center description.
Here, we want to adopt the (according to our reading) same point of view
that analytic properties of polynomial wave functions, while fundamentally related to conformal edge theories,\cite{MR} are not fundamentally essential to the topological order of the state.

Finally, quantum Hall parent Hamiltonians in the second quantized form \eqref{2ndquant}
give rise to frustration free 1D lattice models. Indeed, any zero energy eigenstate of \Eq{2ndquant}
must be a {\em simultaneous} zero energy eigenstate of each of the positive operators
${Q_R^m}^\dagger Q_R^m$. Equivalently, the state must be annihilated by each of the operators
$Q_R^m$, i.e., it must satisfy the zero mode condition
\be
 Q_R^m\, \ket{\psi}=0\;\;\mbox{for all $R$, $m$ included in \Eq{1stquant}.}
\ee
There has been much interest in general properties of such frustration free lattice models recently,
in both 1D and in higher dimensions,\cite{AKLT, nachtergaele, Kennedy92, Bravyi09, NdB, Yoshida, CJKWZ, MZ, schuch, Bravyi, Darmawan} especially, in connection with matrix-product like ground states such models may have.
In particular, for the cylinder and torus geometries, the operators $Q_R^m$ are related by lattice translations. Hence for these geometries in particular, the model has much in common with other frustration free 1D lattice models that arise in solid state, e.g. magnetic, context.
However, the models of the form \eqref{2ndquant} are arguably harder to study.
For frustration free models, the problem is to find the common ground state of all the local terms
entering the full Hamiltonian, while the individual ground state space of {\em one} such local term is
typically easy to characterize. For \Eq{2ndquant}, already finding the ground state subspace
of one operator ${Q_R^m}^\dagger Q_R^m$ is a highly non trivial task, owing to the
exponentially decaying but non-local character of each such term.
This problem was solved in Ref. \onlinecite{ortiz}, for a general class of models of this type,
by making contact with the integrable structure of the hyperbolic Richardson-Gaudin model.
Here, however, the focus will be entirely on finding zero modes of the full Hamiltonian \eqref{2ndquant}.

For the reasons given above, we'd like to have strategies to treat Hamiltonians
that are {\em given} in the form \eqref{2ndquant}, detached from the context of Landau levels and the analytic structure of their first quantized wave functions. To the best of our knowledge, such strategies are currently lacking, despite the recently appreciated matrix product structure of the (Laughlin) ground states of \Eq{1stquant}.\cite{dubail, zaletel}
We will restrict ourselves to the case where the coefficients $\eta_{R,x}^m$ correspond to
Haldane pseudo-potentials in the disk, sphere, or cylinder geometry.
Then by construction, the Hamiltonian \Eq{2ndquant} is frustration free, i.e., there are states
satisfying
the zero mode condition \Eq{zeromode}.
The question we wish to answer is how this fact can be understood in terms
of algebraic properties of the operators $Q_R^m$, ${Q_R^m}^\dagger$.
This is indeed far from obvious.
For the model at hand, we could of course ``go back'' by making connection
with the language of first quantized analytic Landau level wave functions.
However, this would certainly preclude an understanding of the zero mode property
in a manner that is more intrinsic to the second quantized form in which the model
is presented in \Eq{2ndquant}.
Moreover, this approach would also require a large amount of ingenuity
if we did not already know how to re-cast the model in its original first-quantized form.
Indeed, in going from analytic wave functions
to the second quantized ``lattice'' description, information has been dropped about
the dynamical momenta that determine the structure of a Landau level.
As discussed initially, after Landau level projection one is working in
a Hilbert space ${\cal H}_\omega$ that is ``guiding center only''.
In contrast, the original analytic wave functions live in a larger Hilbert space
$\cal H$, which is isomorphic to ${\cal H}_\pi \otimes {\cal H}_\omega$,
where ${\cal H}_\pi$ is associated with the dynamical momenta of the system.
(See also Ref. \onlinecite{RR11} for a recent discussion).
It is only when the embedding
\be
  {\cal H}_\omega \hookrightarrow {\cal H}\,,
\ee
which in principle can be done in infinitely many ways, is defined in exactly the
right manner that we recover the analytical properties of the Laughlin state
that made the model tractable to begin with.\cite{Haldane11}
Here, we do not wish to ``look back'' at the larger Hilbert space
${\cal H}$, but instead take on the model as given in \Eq{2ndquant},
and find a way to understand its frustration free character
in a manner that is intrinsic to the algebraic properties
of the operators $Q_R^m$ in terms of which it is defined there.
It is our hope that this approach will eventually pave the road to an
even larger class of frustration free lattice models.

%The remainder of this paper is organized as follows ...

\section{Algebraic treatment of zero modes}

\subsection{General properties}

In this sub-section only, we will consider the general class of Hamiltonians
given by \eqref{2ndquant}, with interaction parameters $\eta^m_{R,x}$ not necessarily
identical to those obtained from Haldane pseudo-potentials.
For such general Hamiltonians, many useful properties of zero modes
are known from Ref. \onlinecite{ortiz}, under the proviso that such zero modes
exist, i.e., that the Hamiltonian is frustration free (up to some filling factor).
To state these properties, let us first make the mathematical setup more precise.
First, we consider the Landau level ``lattice'' space as half-infinite,
as it is natural to the disk geometry or that of a half-infinite cylinder.
That is, orbitals created by the operators  $c_r^\dagger$ are labeled by a
non-negative integer $r$, and we will write all equations with the convention
\be\label{crzero}
   c_r=c_r^\dagger\equiv 0 \;\;\text{for } r<0
\ee
in mind.
We note that the zero modes we consider will generally occupy only a finite range of orbitals,
and thus remain zero modes whenever a sufficiently large cutoff in orbital space is introduced,
where only orbitals below this cutoff are retained. Thus, while we will not explicitly work with such a cutoff, all of the following is equally relevant to the spherical geometry, where the Hilbert space dimension is generally finite. To this end, we note that each state will be characterized
by a particle number $N$, and a ``maximum occupied orbital'' $r_{\sf max}$, where
\be
  r_{\sf max} =\text{max} \{r | \langle\psi | c^\dagger_r c_r|\psi\rangle \neq 0\}\,,
\ee
and we always leave the $\psi$-dependence of $N$ and $r_{\sf max}$ implicit.
We then define the filling factor as
\be\label{ff}
 \nu = \frac{N-1}{r_{\sf max}}\,,
\ee
where the $-1$ in the numerator takes into account the topological shift for Laughlin states.
We now introduce $f=0$ ($f=1$) for bosons (fermions) and assume that $(-1)^f=(-1)^M$ in \Eq{2ndquant}. The symbols $\eta^m_{R,x}$ are expected to have the symmetry
$\eta^m_{R,-x}=(-1)^f \eta^m_{R,x}$. Note that the sum over $m$ in \Eq{2ndquant}
runs over $D=(M-f)/2$ terms. We then define $M(R)$ as the $D\times D$ matrix
\be
\arraycolsep=1.4pt\def\arraystretch{2.2}
   M(R)_{ij=0\dotsc D-1}= \left\{ \begin{array}{lr}
        \eta^{2j+f}_{R,i+f }  & \text{for } 2R\;\text{even}\\
         \eta^{2j+f}_{R,i+\frac{1}{2}}  &  \text{for } 2R\;\text{odd}\,,
   \end{array} \right .
\ee
where we use the convention
\be\label{etacond}
  \eta^m_{R,x}=0 \;\;\text{for}\; R<|x|,
\ee
setting to zero all coefficients that act on unphysical orbitals with negative index.
Thus, for given $R$, the matrix $M(R)$ contains the parameters determining the interaction
at the $D$ closest distances.
Then, under the general condition that for all $R=0,\frac 12,1\dotsc$, the matrix
$M(R)$ has the maximum rank possible given the constraint \eqref{etacond},
the results of Ref. \onlinecite{ortiz} give the following:

{\bf Theorem 1.a}
The Hamiltonian \eqref{2ndquant} has no zero modes with filling factor $\nu>1/M$.

{\bf Theorem 1.b}
If the Hamiltonian \eqref{2ndquant} has a zero mode at filling factor $\nu=1/M$,
it is unique.

Both of these theorems are direct consequences of the following. We will say
that an occupation number eigenstate $|\{n\}\rangle=\ket{n_0,n_1,\dotsc}$
satisfies the ``$M$-Pauli principle'', in the sense of Ref. \onlinecite{BH2}, if there is no more than $1$ particle in any $M$ consecutive orbitals. We define ``inward-squeezing'' operations\cite{BH2} of the form
\be \label{inward}
     c^\dagger_{j}c^\dagger_{i} c_{i-d} c_{j+d}
\ee
where $i \leq j$ and  $d>0$. Then, we say that an occupation number
eigenstate $\ket{\{n_i\}}$ can be obtained from an occupation number eigenstate  $\ket{\{n'_i\}}$
by inward-squeezing
 if $\ket{\{n_i\}}$ can be obtained from  $\ket{\{n'_i\}}$ (up to some normalization and phase)
 by repeated application of operations of the form \eqref{inward} (i.e., center-of-mass conserving
 inward pair hopping processes).  We can expand any given state $\ket{\psi}$ in occupation
 number eigenstates:
 \be\label{expansion}
     \ket{\psi}= \sum_{\{n_i\}} C_{\{n_i\}} \ket{\{n_i\}}\,.
 \ee

 Then we have the following\cite{ortiz}

 {\bf Theorem 1}
If $\ket{\psi}$ is a zero mode of the Hamiltonian \eqref{2ndquant}, and the corresponding matrices
$M(R)$ satisfy the maximum rank criterion defined above, then any basis state
$\ket{\{n_i\}}$ appearing in \eqref{expansion} with $C_{\{n_i\}}\neq 0$ can be obtained from
a $\ket{\{n'_i\}}$ (depending on $\ket{\{n_i\}}$ in general) through inward squeezing, where  $C_{\{n'_i\}}\neq 0$ and $\ket{\{n'_i\}}$ satisfies the $M$-Pauli principle.

It is then easy to see that Theorems 1.a and 1.b follow, respectively, from the observations that
there is no $\ket{\{n_i\}}$ satisfying the $M$-Pauli principle at filling factor $>\frac{1}{M}$,
and exactly one such at filling factor $\frac{1}{M}$. The latter is the well-known thin torus
limit\cite{TT, RH,  BK1, seidel05, SL,seidellee07, BK2, BK, SY08, ABKW, BHHKV, seidel10, LBSH, SY11, papic, WS} or root partition\cite{BH1, BH2, BH3,  regnault, ortiz} of the $\frac{1}{M}$-Laughlin state,
$10\dotsc 010\dotsc 010\dotsc01$, where $1$s are separated by exactly $M-1$ zeros.

We may also remark the following trivial observation:

{\bf Proposition 2}
 If the Hamiltonian \eqref{2ndquant} has zero modes at some filling factor $\nu^\ast$ and
  with particle number $N>1$,
 then there are also zero modes at filling factors $\nu<\nu^\ast$.

 This simply follows from the zero mode condition \eqref{zeromode}, together
 with the observation that $[Q_R^m, c_r]=0$. Hence we can always generate
 new zero modes from old ones by acting with destruction operators $c_r$.
 In general, however, we may not hope to generate all possible zero modes
 in this way. For the special case of the Laughlin-state parent Hamiltonians only, a more complete prescription using second quantization was given
 in Ref. \onlinecite{ortiz}, where it was noted that certain
particle number conserving operators generate new zero modes at higher $r_{\sf max}$
when acting on a given zero mode $\ket{\psi}$.\footnote{This corresponds to the familiar multiplication with symmetric polynomials in first quantized language.}
A variant of these operators will be defined below.
% also generate zero modes at lower filling factor when applied to a given zero mode $\ket{\psi}$.
% An alternative set of generators will be given below.

 It is worth emphasizing again that while all facts stated in this section are known
 for the Laughlin-state parent Hamiltonians from first quantized wave function
 considerations\cite{RH} and known squeezing properties
 of ``special'' wave functions such as Laughlin states,\cite{RH}
 all of the above was shown in Ref. \onlinecite{ortiz} for zero modes of the more general
 class of Hamiltonians \eqref{2ndquant}, under the general maximum rank condition
 stated above. This maximum rank condition is easily adapted to other generalized
 Pauli principles and $n$-body operators. None of this makes use of analytic clustering
properties (which are in fact not necessary, as, e.g., demonstrated by the examples given
in Refs. \onlinecite{Nakamura}). Given the above,
it seems that most known facts
about Laughlin-state parent Hamiltonians are already within reach of a purely algebraic, or second-quantized derivation. There is, however, a key ingredient thus far missing:
 Namely, the fact that there exists, to begin with, a
zero mode at the special ``incompressible'' filling factor $\frac{1}{M}$.
Once this is established, further zero modes can be generated using the operators defined in
Ref. \onlinecite{ortiz}, or the operators given in \Eq{en} below.
To understand the existence of a special zero mode at filling factor $\frac 1 M$ (whose uniqueness is then guarantied, e.g.,  by Theorem 1.b) in terms of algebraic properties of the operators $Q^m_R$ is the main goal of this paper.

\subsection{Recursive definition of the Laughlin state in second quantization}

 The Laughlin state at filling factor $\nu=\frac{1} {M}$ and its zero modes,
 which for $\nu<\frac{1}{M}$ physically represent quasi-hole and edge excitations,
 can be characterized as forming the common null space of the operators
 $Q^m_R$ for all $R=0, \frac {1}{2}, 1, \frac{3}{2}\dotsc$
 and all $m=M-2, M-4\dotsc 0$ (1) for bosons (fermions).
 As explained in Ref. \onlinecite{ortiz}, the form factors
 $\eta^m_{R,x}$ may be taken to be of the form
 \be\label{eta}
 \eta^m_{x}=x^m.
\ee
 Formally, this corresponds to working with zero modes in the
 cylinder geometry in the limit where the cylinder radius goes to infinity.
 The zero modes in the disk geometry, that of any cylinder of finite thickness,
 or the sphere are in one-to-one correspondence with the zero modes obtained in this way,\footnote{See. Ref. \onlinecite{ortiz} for a second-quantized proof.}
 where for the sphere, an upper cutoff in $r_{\sf max}$ must be introduced.
 We also work with a lower cutoff in orbital space, corresponding to a half-infinite geometry,
 imposed by the condition \eqref{crzero}.
 This condition will always be left understood in the following.
 We note that for a general cylinder with radius $R_y\equiv1/\kappa$,
 the coefficient $\eta^m_x$ corresponding to the $m$th Haldane pseudo-potential
 is given by:\cite{ortiz}
 \be\label{hermite}
   \eta^m_x= {\cal N}_m \sqrt{\kappa} \,H_{m}(\sqrt{2}\kappa x) e^{-\kappa^2 x^2},
 \ee
 where ${\cal N}_m$ is a normalization factor, and $H_m$
 is the $m$th Hermite polynomial.
 The simple limiting form   \eqref{eta} is justified by first taking linear
 combinations in $m$ of the $Q^m_R$ at {\em finite} $\kappa$
  such that the polynomials in \eqref{hermite} are replaced by monomials,
  and then taking the limit $\kappa\rightarrow 0$.
  We note that forming such linearly independent new linear combinations does not affect the common
  null space of the operators $Q^m_R$.

 As explained in the preceding paragraph, we now focus on the question
 of the existence of a zero mode at filling factor $\nu=\frac{1}{M}$
 satisfying
 \be\label{zeromode}
 \begin{split}
     Q^m_R |\psi_N\rangle=0 \quad\text{for all $R$, and for $0\leq m < M$}\\
     m \equiv M \mod 2.
 \end{split}
 \ee
 Here, the subscript $N$ stands for an $N$-particle state with $r_{\sf max}= M(N-1)$,
 whose existence we will prove inductively.
 We also introduce the ``angular momentum'' operator
 \be
 L=\sum_r rc^\dagger_r c_r,
 \ee
 and anticipate that $\ket{\psi_N}$ will be an $L$-eigenstate
 with eigenvalue $L=\frac 12 MN(N-1)$, as befits a $\nu=\frac 1 M$
 Laughlin state.

 We seek a recursive definition for $\ket{\psi_N}$ for which we can prove
 the zero mode property inductively. Our general strategy will be the following.
 We start with the trivial identity
 \be
     \ket{\psi_N}=\frac 1 N \sum_r c^\dagger_r c_r \ket{\psi_N}\,.
 \ee
 We have observed above already that if $\ket{\psi_N}$ is an $N$-particle zero mode,
 then $c_r\ket{\psi_N}$ is a zero mode with $N-1$ particles.
 As such, it can be generated from the $N-1$ particle incompressible Laughlin state
 at $\nu=\frac 1 M$ through the application of an appropriate particle number {\em conserving} operator that creates
 zero energy edge excitations.
 We thus conjecture that there is a well-defined operator
 $P_\ell$, creating an edge excitation that increases the angular moment
 $L$ by $\ell$ units while conserving particle number, such that
 \be \label{cr}
       c_r\ket{\psi_N} = P_{M(N-1)-r} \ket{\psi_{N-1}}
 \ee
 holds. Here, we have used the fact that the
 incompressible Laughlin state at filling factor $\frac 1 M$ has
 $L$-eigenvalue $\frac 12 M N (N-1)$.
 This leads to a recursive definition of the $N$-particle
 Laughlin state in terms of the  $N-1$-particle Laughlin state:
 \be\label{recursive}
     \ket{\psi_N} = \frac{1}{N} \sum_{r\geq 0} c_r^\dagger P_{M(N-1)-r} \ket{\psi_{N-1}}\,.
 \ee
 The strategy will hence be to identify the precise form of the operator $P_\ell$, and prove inductively
 that indeed \Eq{recursive} defines an $N$-particle zero mode at filling factor $1/M$, as long as $\ket{\psi_{N-1}}$ has the same property for $N-1$ particles.
 We note that the strategy given here is quite general, and the same logic would in principle lead to a recursive expression similar to \Eq{recursive} for other types of quantum Hall states.
 We focus on the case of the Laughlin state here.
In this special case, the relation \eqref{recursive} we seek turns out to be a second quantized
rendering of Read's recursive formula for the Laughlin state using the ``string order parameter''.\cite{readOP} The relation of our formalism to the string order parameter will be explored
by work in parallel,\cite{mazaheri_et_al} and will not be elaborated further in the following.

\subsection{The generators of edge excitations and relevant commutation relations\label{algebra}}

As motivated above, we are interested in particle number { conserving} operators that
generate new zero modes when acting on given zero modes. We will think of such
operators as generators of (zero energy) edge excitations.

Consider the operators
\be
    e_n=\frac{1}{n!} \sum_{i_1\geq0,i_2\geq 0,\dots i_n\geq 0} c^\dagger_{i_1+1}c^\dagger_{i_2+1}\dotsc c^\dagger_{i_n+1} c_{i_n}\dotsc c_{i_2}c_{i_1}\,,\label{en}
\ee
where the operators $c_r$, $c_r^\dagger$ satisfy standard bosonic (fermionic) commutation (anticommutation) relations for $M$ even (odd), and we fix an integer $M>1$ here and in the following. We also define $e_0=\mathbb{1}$ and $e_n=0$ for $n<0$. $e_n$ can increase the angular momentum of each of $n$ orbitals out of occupied orbitals by 1. The action of $e_n$ acting on zero modes corresponds to multiplying the Laughlin wave function with an elementary symmetric polynomial in the first quantized language.
These operators conserve particle number,
and have the property that if $\ket{\psi}$ is a zero mode, then so is $e_n \ket{\psi}$.
This follows since by definition, a zero mode is annihilated by all the operators
$Q^m_R$, with $m$ and $R$ as in \Eq{zeromode}, and we have the commutator

\begin{equation}
\begin{split}\label{Q & e}
[Q_R^m,{e_n}]=
&{e_{n - 2}}Q_{R - 1}^m + \\
& \hspace*{-0.0cm}e_{n - 1} \sum_{\substack{0\leq k\leq m-f\\ (-1)^k=1}}{m \choose k}{{2^{1 - k}}Q_{R - 1/2}^{m - k}}\,,
\end{split}
\end{equation}
which vanishes on all zero modes.
The operators $e_n$ are not the same edge mode generators as those defined in Ref. \onlinecite{ortiz}.
The relation between the latter and the $e_n$ is not of any importance in the following, and will be
clarified
by work in parallel.\cite{mazaheri_et_al}
In terms of the $e_n$, we now define new operators
\be
 P_\ell =(-1)^\ell \sum_{n_1+n_2+\dots +n_M=\ell}e_{n_1} e_{n_2}...e_{n_M}.\label{Pl}
\ee
The latter are likewise particle number conserving generators of zero modes,
since the $e_n$ have this property. These operators $P_\ell$ depend on $M$ as seen from the definition, but we leave the dependence implicit. It is also understood that $P_0=\mathbb{1}$ and $P_l=0$ for $l<0$. It is furthermore easy to see that $P_\ell$ raises the
angular momentum by $\ell$.
We now {\em define} the $\nu=\frac 1 M$ Laughlin state through \Eq{recursive},
where
\be\label{start}
   \ket{\psi_{N=0}}=\ket{0}\,,
\ee
$\ket{0}$ being the vacuum, which also leads to $\ket{\psi_{N=1}}=c^\dagger_0 \ket{0}$.
For the time being, we {\em assert} that \Eq{cr} follows from this definition and
from \Eq{Pl}. It turns out that most technical difficulties can be attributed to the proof of
this assertion, which we relegate to Sec.~\ref{Pproof}. Our key result, namely,
that \Eq{recursive} defines a zero mode at filling factor $\frac 1 M$, follows rather
easily from \Eq{cr} and the following commutation relations:
\be
   [Q_R^m, c_r]=0,
\ee

\be\label{Qandcr+}
   [Q_R^m, c^\dagger_r]=2(-1)^f (R-r)^m c_{2R-r},
\ee

\be\label{Pandcr}
   [c_r, P_l]=\sum_{\substack{1\leq k\leq M}}(-1)^k{M \choose k}P_{l-k} c_{r-k},
\ee
\be\label{Pandcr+}
   [P_l, c^\dagger_r]=\sum_{\substack{1\leq k\leq M}}(-1)^k{M \choose k} c^\dagger_{r+k}P_{l-k},
\ee
and \be[e_m, e_n]=0.\ee
Another useful way to write \Eq{Pandcr} is

\be\label{Pandcr2}
   c_r P_l=\sum_{\substack{0\leq k\leq M}}(-1)^k{M \choose k}P_{l-k} c_{r-k},
\ee
and similarly for \Eq{Pandcr+}.

\subsection{Proof of zero mode property of $\ket{\psi_N}$}

We proceed by showing that given all of the above, and assuming \Eq{cr} to be true
for now, it follows that Eqs. \eqref{recursive}, \eqref{start} define a non-vanishing zero
mode at the ``incompressible'' filling factor $1/M$.

To see that \Eq{recursive} gives a zero mode, we proceed  inductively, starting with $N=2$:

\[
\begin{split}
& \ket{\psi_{N=2}}  \\
& = \frac{1}{2} \sum_{r\geq 0} c_r^\dagger P_{M-r} c^\dagger_0 \ket{0} \\
& =\frac{1}{2} \sum_{r\geq 0} c_r^\dagger \sum_{\substack{0\leq k\leq M}}(-1)^k{M \choose k} c^\dagger_{k}P_{M-r-k}\ket{0}\,,
\end{split}
\]
where the last line follows from \Eq{Pandcr+}. For positive index, $P_{M-r-k}$ is the sum of products of $e$ operators that have annihilation operators on the right, thus $P_{M-r-k}\ket{0}$ gives zero
unless $M-r-k=0$.
%when $M-r-k>0$. So is the case when $M-r-k<0$ because $P_{M-r-k}=0$ for $M-r-k<0$.
Therefore

\be \label{psain2}
\ket{\psi_{N=2}}=\frac{1}{2} \sum_{r\geq 0}(-1)^{M-r}{M \choose r} c_r^\dagger  c^\dagger_{M-r}\ket{0}.
\ee
It is easy to show that \be
\begin{split}
&
Q^m_R
\ket{\psi_{N=2}}\\
&
=\delta_{M,2R}(-1)^{3R}\sum_{\substack{-R\leq x\leq R}}x^m(-1)^{x}{2R \choose R+x} \ket{0}\\
&=\delta_{M,2R}(-1)^{M}\sum_{\substack{0\leq x'\leq M}} {\left( {x' - \frac{M}{2}} \right)^m}(-1)^{x'}{M \choose x'} \ket{0}\\
&=0\,,
\end{split}
\ee
where the sum in the last line
follows from the fact that
\footnote{
Let $[j]_0=1$, $[j]_m=j(j-1)\dotsc (j-m+1)$ for $m>0$.
Then $\sum_{{0\leq j\leq M}}  [j]_m (-1)^j  {M \choose j}=\sum_{{m\leq j\leq M}}  \frac{j!}{(j-m)!}  (-1)^j {M \choose j} =(-1)^m [M]_m (1-1)^{M-m} \,.$ This gives $0$ for $0\leq m < M$. Since $[j]_m$ is clearly an $m$th degree polynomial in $j$,
we can make new linear combinations of the latter identities to obtain \Eq{jm}. Extending consideration to $m=M$ gives all the ingredients for the interesting identity \cite{Ruiz96}
$\sum_{j=0}^M (-1)^j{M\choose j}(x-j)^M= M! \quad \forall x$.}

 \cite{Ruiz96}

\be\label{jm}
\sum_{\substack{0\leq j\leq M}}  j^m (-1)^j  {M \choose j} = 0\quad \text{for } 0\leq m < M\,.
\ee
Thus\be Q^m_R
\ket{\psi_{N=2}}=0. \ee
Now assume
\be\label{zero}
 \begin{split}
     Q^m_R |\psi_{N-1}\rangle=0 \quad\text{for all $R$, and for $0\leq m < M$}\\
     m \equiv M \mod 2\,.
 \end{split}
 \ee
We have
\[
\begin{split}
& Q^m_R|\psi_N\rangle \\
& =\frac{1}{N} \sum_{r\geq 0} (c_r^\dagger Q^m_R+2(-1)^f (R-r)^m c_{2R-r}) P_{M(N-1)-r} \ket{\psi_{N-1}} \\
& =\frac{1}{N} \sum_{r\geq 0} 2(-1)^f (R-r)^m c_{2R-r} P_{M(N-1)-r} \ket{\psi_{N-1}}\\
& =\frac{1}{N} \sum_{r\geq 0} 2(-1)^f (R-r)^m c_{2R-r} c_{r} \ket{\psi_{N}}\\
& =\frac{2}{N} Q^m_R \ket{\psi_{N}}\,,
\end{split}
\]
where we have used \Eq{recursive} and \Eq{Qandcr+} to get the second line. The third line uses the fact that $P_{M(N-1)-r} \ket{\psi_{N-1}}$ also satisfies the zero mode condition \eqref{zero}, since $P_\ell$, being a product of $e$-operators, generates new zero modes from old ones. The fourth line follows from \Eq{cr}. Therefore, for $N  \geq 3$, if $\ket{\psi_{N-1}}$ satisfies the zero mode condition, so will $\ket{\psi_{N}}$. Finally,  $\ket{\psi_{N=0}}$ and $\ket{\psi_{N=1}}$ are trivially zero modes. Thus, all $\ket{\psi_{N}}$ satisfy the zero mode property \eqref{zeromode}.
We will still need to demonstrate that $\ket{\psi_{N}}$ has filling factor $1/M$, and in particular
does not vanish for any $N$. Before doing so in Sec.~ \ref{nonzero}, we attend to the
technical heart of the proof, \Eq{cr}.

\subsection{Expressing electron holes through edge excitations}\label{Pproof}

We note that \Eq{cr} expresses an electron hole inserted
into an $N$-particle Laughlin state through a superpositions of general edge excitations created
on top of an $N-1$ particle Laughlin state.
We believe that
 this relation could prove useful in itself beyond
the application given here.
We prove \Eq{cr} inductively:

For $N=1$, \be
c_r\ket{\psi_{N=1}}=P_{-r}\ket{0}
\ee is satisfied for $r=0$, and for $r\neq 0$ both sides vanish identically.
 Now for some $N>1$, we make the assumption that
 \be
c_r\ket{\psi_{N-1}}=P_{M(N-2)-r}\ket{\psi_{N-2}},
\ee
The definition of $\ket{\psi_N}$, \Eq{recursive},
 gives

\begin{equation}
\begin{split}\label{crpsai}
c_r\ket{\psi_{N}}=
&c_r\frac{1}{N} \sum_{r'\geq 0} c_{r'}^\dagger P_{M(N-1)-r'} \ket{\psi_{N-1}}\\
& =\frac{1}{N}\sum_{r'\geq 0} ({\delta _{r r'}}+(-1)^fc_{r'}^\dagger c_{r}) P_{M(N-1)-r'} \ket{\psi_{N-1}}\,.
\end{split}
\end{equation}
Employing
\Eq{Pandcr2}, the last term of \Eq{crpsai} \be\frac{1}{N}\sum_{r'} (-1)^f c_{r'}^\dagger c_{r} P_{M(N-1)-r'} \ket{\psi_{N-1}}\ee is found to be
\begin{equation}
\begin{split}
&\frac{(-1)^f}{N}\sum_{\substack{0\leq k\leq M-1}} (-1)^k{M \choose k}\sum_{r'} c_{r'}^ \dagger P_{M(N-1)-r'-k}c_{r-k} \ket{\psi_{N-1}}\\
&+\frac{1}{N}\sum_{r'} c_{r'}^\dagger P_{M(N-1)-r'-M}c_{r-M} \ket{\psi_{N-1}},
\end{split}\label{eq2}
\end{equation}
where we split off the last term. We now use the induction assumption,
according to which  $c_{r-M} \ket{\psi_{N-1}}$ is equal to $ P_{M(N-1)-r}\ket{\psi_{N-2}}$.
Therefore the last term of \Eq{eq2} can be further simplified to read \be \frac{1}{N}\sum_{r'} c_{r'}^\dagger P_{M(N-2)-r'}P_{M(N-1)-r}\ket{\psi_{N-2}}. \ee Here, we may now change the order of $P_{M(N-2)-r'}$ and $P_{M(N-1)-r}$, since they are products of commuting $e_n$'s. Then we can use \Eq {Pandcr+} to rewrite \be \frac{1}{N}\sum_{r'} c_{r'}^\dagger P_{M(N-1)-r} P_{M(N-2)-r'}\ket{\psi_{N-2}}\ee
 as

 \begin{equation}
 \begin{split}
 &\frac{1}{N}  P_{M(N-1)-r}\sum_{r'} c_{r'}^\dagger P_{M(N-2)-r'}\ket{\psi_{N-2}}-\frac{1}{N} \sum_{r'}\\
 &\sum_{\substack{1\leq i\leq M}}(-1)^i {M \choose i} c_{r'+i}^\dagger P_{M(N-1)-r-i}P_{M(N-2)-r'}\ket{\psi_{N-2}},
 \end{split}
 \end{equation} where the first term is just $ \frac{N-1}{N}  P_{M(N-1)-r}\ket{\psi_{N-1}}$  utilizing our induction assumption and the second term can be written as
\begin{equation}
  \begin{split}
  &\frac{1}{N} \sum_{\substack{0\leq k\leq M-1}}(-1)^{k+M+1} {M \choose k}\\
  &\sum_{r''} c_{r''}^\dagger P_{M(N-2)-r+k}
  P_{M(N-1)-r''-k} \ket{\psi_{N-2}}
  \end{split}\label{eq3}
  \end{equation}
 after we make changes of variables $r''=r'+i$ and $k=M-i$. \Eq{eq3} is seen to cancel the first term of \Eq{eq2} after we change the order of two $P$ operators and use once more our induction assumption. Finally we get $c_r\ket{\psi_{N}}=(\frac{1}{N}+\frac{N-1}{N}) P_{M(N-1)-r}\ket{\psi_{N-1}}=P_{M(N-1)-r}\ket{\psi_{N-1}}$, thus completing our induction to prove \Eq{cr}.

\subsection{Properties of $\ket{\psi_N}$}\label{nonzero}

In the above, we have shown that the recursively defined state \eqref{recursive}
has the zero mode property for all $N$. The proof was based solely on the algebraic
properties described in Sec. \ref{algebra}. To achieve our initial goal, we must also demonstrate
that $\ket{\psi_N}$ is a {\em non-vanishing} $N$-particle state at filling factor $1/M$.

To this end, we define the ``thin cylinder state'' $\ket{\tilde\psi_N}$ discussed below Theorem 1 via
\be
\ket{\tilde\psi_N}=\ket{10\dotsc010 \dotsc 010\dotsc},
\ee
where exactly $N$ $1$'s are separated by sequences of $M-1$ zeros.
We then assert the following

{\bf Proposition 3} The state $\ket{\psi_N}$ defined by \Eq{recursive} is dominated by the basis state $\ket{\tilde\psi_N}$
with $\langle \tilde \psi_N|\psi_N\rangle\neq 0$.

Here, the notion of dominance means, as usual,\cite{BH2, BH3},
that all basis states appearing in the expansion \eqref{expansion} can be obtained
from $\ket{\tilde\psi_N}$ via inward squeezing operations, as explained  following
\Eq{inward}.
The proposition in particular implies all the desired information about $\ket{\psi_N}$.
It clearly implies that $\ket{\psi_N}$ is non-zero. It is also easy to see, given the definition \eqref{ff}, that any state dominated by $\ket{\tilde\psi_N}$ has a filling factor of at most $1/M$, and has precisely filling factor $1/M$ if $\langle \tilde \psi_N|\psi_N\rangle\neq 0$.

  Again, we prove Proposition 3 inductively.
 For $N=1$, the statement is obvious.
 Assuming that Proposition 3 has been proven for $N-1$ with $N\geq 2$, we consider
 $\ket{\psi_N}$ as defined through \Eq{recursive}.
 It is elementary to see from this equation that if $\ket{\psi_{N-1}}$ is dominated
 by the basis state     $\ket{\tilde\psi_{N-1}}$, then $\ket{\psi_N}$ must have {\em at least}
 filling factor $1/M$, i.e., its $r_{\sf max}$ can be at most $M(N-1)$.
 On the other hand, by Theorem 1.a, since we know that $\ket{\psi_N}$ has the zero mode property,
 it must have a filling factor of {\em exactly} $1/M$, so long as it is non-zero.
 If indeed $\ket{\psi_N}$ is non-zero, by Theorem 1 its expansion \eqref{expansion} into occupation number
 eigenstates must then also be dominated by $\ket{\tilde \psi_N}$, the latter being the only
 such state that satisfies the $M$-Pauli principle at filling factor $1/M$.
 Therefore, all that remains to show is that $\langle \tilde \psi_N|\psi_N\rangle\neq 0$.

By the induction assumption, we may write
\be\label{psiN1}
    \ket{\psi_{N-1}} = C_{\tilde\psi_{N-1}} \ket{\tilde\psi_{N-1}} +\sum_{\ket{\{n_i\}}\neq \ket{\tilde\psi_{N-1}}}
    C_n \ket{\{n_i\}}
\ee
with $C_{\tilde\psi_{N-1}} \neq 0$, and every $\ket{n}$ appearing in the sum being dominated by
$\ket{\tilde\psi_{N-1}}$. When this is plugged into \Eq{recursive},
one may see that
\begin{equation}
\begin{aligned}
 & \frac{1}{N} \sum_{r\geq 0} c_r^\dagger P_{M(N-1)-r} \ket{\tilde\psi_{N-1}}\\
 =  & \frac{1}{N}\sum_{r\geq 0} c_r^\dagger P_{M(N-1)-r} c^\dagger_{0}c^\dagger_{M} \dots c^\dagger_{(N-2)M }\ket{0}\\
 =& \frac{1}{N} \sum_{0\leq k_0,k_1,\dots , k_{N-2}\leq M }(-1)^{k_0+k_1+\dotso +k_{N-2}}{M \choose k_0}\dotso {M \choose k_{N-2}} \\
 & c^\dagger_{(N-1)M-k_0-\dotso-k_{N-2}}  c^\dagger_{k_0}c^\dagger_{k_1+M} \dots c^\dagger_{k_{N-2}+(N-2)M }\ket{0} \,
\end{aligned}
\end{equation}
after using the same approach leading to \Eq{psain2}.
It is clear now why  $r_{\sf max}$ can be at most $M(N-1)$. To generate $\ket{\tilde\psi_N}$, $ (N-1)M-k_0-\dots-k_{N-2}$ should be equal to $mM$ where $0\leq m\leq N-1$ and $k_{0},  k_{1}+M, \dotso, k_{N-2}+(N-2)M$ each should assume one of the values $0, M, \dotso, (m-1)M, (m+1)M, \dotso, (N-1)M $.

If $m<N-1$, the only index that can assume the value $(N-1)M$ is $k_{N-2}+(N-2)M$, and
this fixed $k_{N-2}=M$. Working our way down in $j$ from $j=N-2$, we find from the same reasoning that $k_j=M$ for $j\geq m$. Then, in order for the first index $(N-1)M-k_0-\dots-k_{N-2}$ to equal
$mM$, all the remaining $k_j$ for $0\leq j<m$ must vanish.

The only solution for given $m$ is thus $ k_0=k_1=\dots= k_{m-1}=0$ and $ k_{m}=k_{m+1}=\dots= k_{N-2}=M$.
Therefore $\ket{\tilde\psi_N}$ is generated from $\ket{\tilde\psi_{N-1}}$ through $N$ possible choices of $m$ all leading to the same coefficient of $\ket{\tilde\psi_N}$, which is $(-1)^{(N-1)M}$. Furthermore, the states dominated by $\ket{\tilde\psi_{N-1}}$ cannot generate $\ket{\tilde\psi_{N}}$. To see this, we act $\frac{1}{N}\sum_{r\geq 0} c_r^\dagger P_{M(N-1)-r}$ on one of those states:
\begin{equation}\label{subdom}
\begin{aligned}
 & \frac{1}{N} \sum_{r\geq 0} c_r^\dagger P_{M(N-1)-r} \ket{n}\\
 =  & \frac{1}{N}\sum_{r\geq 0} c_r^\dagger P_{M(N-1)-r} c^\dagger_{r_0}c^\dagger_{r_1} \dots c^\dagger_{r_{N-2}}\ket{0}\\
 =& \frac{1}{N} \sum_{0\leq k_0,k_1,\dots , k_{N-2}\leq M }(-1)^{k_0+k_1+\dotso +k_{N-2}}{M \choose k_0}\dotso {M \choose k_{N-2}}\\
 & c^\dagger_{(N-1)M-k_0-\dotso-k_{N-2}}  c^\dagger_{k_0+r_0}c^\dagger_{k_1+r_1} \dots c^\dagger_{k_{N-2}+r_{N-2} }\ket{0} \,.
\end{aligned}
\end{equation}
We may assume the $r_j$ to be in ascending order.
$\ket{\{n_i\}}$ is obtained from $\ket{\tilde\psi_{N-1}}$ through inward squeezing defined in \Eq{inward}. The largest index $j_0$ for which $r_{j}$
differs from $Mj$ must then have  $r_{j_0}<Mj_0$.
We may now ask if \Eq{subdom} could make contributions to $\ket{\tilde\psi_N}$.
We can follow the same logic as above, fixing the first index to be $mM$, and then
fixing the $k_j$, starting with $j=N-2$ and working our way down. We will always
find a contradiction once we reach $j_0$, using $r_{j_0}<Mj_0$.

 In all, we find $C_{\tilde\psi_{N}}=(-1)^{(N-1)M}C_{\tilde\psi_{N-1}} $, which does not vanish. This completes the argument that $\langle \tilde \psi_N|\psi_N\rangle\neq 0$.

\section{Discussion and Conclusion}

In this paper, we study a particular class of frustration free lattice Hamiltonians
that arises in the fractional quantum Hall effect via Haldane's pseudo-potential
formalism. There is a great wealth of knowledge on general properties
of frustration free Hamiltonians, which recently was further expanded by
interest in matrix product states. In general, the interrelation between matrix
product states on a lattice and their frustration free parent Hamiltonians is well
understood.\cite{nachtergaele} This, however, is arguably different for the class
we have studied here, which differs from most examples in the literature in that the interaction is
not strictly local.
Despite the fact that the matrix product structure of the Laughlin state has recently
been appreciated,\cite{cirac, dubail, zaletel} to the best of our knowledge,
the only way to understand the frustration free character of its ``lattice'' parent
Hamiltonians involved going back to first quantization -- using a language
of analytic wave functions where the ``lattice character'' is lost,
as we reviewed initially. While the first quantized
language of polynomial wave functions is powerful and has thus far been a primary driving
factor in this field, it is a priori not clear why the inclusion of additional degrees
of freedom (dynamical momenta) is necessary to solve the problem of studying the
zero modes of the lattice Hamiltonians that, as we did, one may choose to regard as the starting point.
Such a point of view is natural and may be of great value especially in the study of parent Hamiltonians for fractional Chern insulators.\cite{Tang11,Sun11,Neu11,Sheng11,Wang11,Regnault11,Ran11,Bernevig12,WangPRL12,WangPRB12,Wu12,Laeuchli12,Bergholtz12,Sarma12,Liu13,chen13, scaffidi}
In this work, we have demonstrated that the frustration free character of  ``lattice'', or second quantized, Haldane-type pseudo-potential Hamiltonians can be understood directly, without going back to a polynomial language. To this end, we explicitly constructed the $1/M$ Laughlin state
in the lattice basis. This was done by iteratively constructing the $N$-particle Laughlin state from the $N-1$ particle one, in what turned out to be a second quantized form of Read's iterative
formula using the order parameter for Laughlin states.\cite{readOP} We have identified
the proper algebra of lattice operators that allows both construction of the Laughlin state
and statement of the  zero mode condition. Using this algebra alone we have demonstrated that
the ``lattice'' Hamiltonians obtained from first quantized Laughlin state parent Hamiltonians have a (unique)
zero mode at the respective highest filling factor $1/M$. From this very fact the entire zero mode structure can be derived, also by using only the second quantized algebraic setting used here.\cite{ortiz, mazaheri_et_al} We believe that these results deepen our insights into the structure of  fractional quantum Hall and Chern insulator type of parent Hamiltonians, as well as frustration free lattice Hamiltonians in general, and in particular, long ranged ones with matrix product ground states of unbounded bond dimension. Their application to other known as well as possibly novel parent Hamiltonians is left as an interesting direction for the future.

\begin{acknowledgments}
This work has been supported by the National Science
Foundation under NSF Grant No. DMR-1206781 (AS). AS would like to thank Z. Nussinov, G. Ortiz, and X. Tang for insightful
discussion.

\end{acknowledgments}
\bibliography{alg}

\begin{thebibliography}{79}%
\makeatletter
\providecommand \@ifxundefined [1]{%
 \@ifx{#1\undefined}
}%
\providecommand \@ifnum [1]{%
 \ifnum #1\expandafter \@firstoftwo
 \else \expandafter \@secondoftwo
 \fi
}%
\providecommand \@ifx [1]{%
 \ifx #1\expandafter \@firstoftwo
 \else \expandafter \@secondoftwo
 \fi
}%
\providecommand \natexlab [1]{#1}%
\providecommand \enquote  [1]{``#1''}%
\providecommand \bibnamefont  [1]{#1}%
\providecommand \bibfnamefont [1]{#1}%
\providecommand \citenamefont [1]{#1}%
\providecommand \href@noop [0]{\@secondoftwo}%
\providecommand \href [0]{\begingroup \@sanitize@url \@href}%
\providecommand \@href[1]{\@@startlink{#1}\@@href}%
\providecommand \@@href[1]{\endgroup#1\@@endlink}%
\providecommand \@sanitize@url [0]{\catcode `\\12\catcode `\$12\catcode
  `\&12\catcode `\#12\catcode `\^12\catcode `\_12\catcode `\%12\relax}%
\providecommand \@@startlink[1]{}%
\providecommand \@@endlink[0]{}%
\providecommand \url  [0]{\begingroup\@sanitize@url \@url }%
\providecommand \@url [1]{\endgroup\@href {#1}{\urlprefix }}%
\providecommand \urlprefix  [0]{URL }%
\providecommand \Eprint [0]{\href }%
\providecommand \doibase [0]{http://dx.doi.org/}%
\providecommand \selectlanguage [0]{\@gobble}%
\providecommand \bibinfo  [0]{\@secondoftwo}%
\providecommand \bibfield  [0]{\@secondoftwo}%
\providecommand \translation [1]{[#1]}%
\providecommand \BibitemOpen [0]{}%
\providecommand \bibitemStop [0]{}%
\providecommand \bibitemNoStop [0]{.\EOS\space}%
\providecommand \EOS [0]{\spacefactor3000\relax}%
\providecommand \BibitemShut  [1]{\csname bibitem#1\endcsname}%
\let\auto@bib@innerbib\@empty
%</preamble>
\bibitem [{\citenamefont
  {Schr\ifmmode\ddot{o}\else\"{o}\fi{}dinger}(1926)}]{schrodinger}%
  \BibitemOpen
  \bibfield  {author} {\bibinfo {author} {\bibfnamefont {E.}~\bibnamefont
  {Schr\ifmmode\ddot{o}\else\"{o}\fi{}dinger}},\ }\href {\doibase
  10.1002/andp.19263840602} {\bibfield  {journal} {\bibinfo  {journal} {Annalen
  der Physik}\ }\textbf {\bibinfo {volume} {384}},\ \bibinfo {pages} {489}
  (\bibinfo {year} {1926})}\BibitemShut {NoStop}%
\bibitem [{\citenamefont {Pauli}(1926)}]{alg_hydrogen}%
  \BibitemOpen
  \bibfield  {author} {\bibinfo {author} {\bibfnamefont {W.}~\bibnamefont
  {Pauli}},\ }\href {\doibase 10.1007/BF01450175} {\bibfield  {journal}
  {\bibinfo  {journal} {Z. Phys.}\ }\textbf {\bibinfo {volume} {36}},\ \bibinfo
  {pages} {336} (\bibinfo {year} {1926})}\BibitemShut {NoStop}%
\bibitem [{\citenamefont {Haldane}(1983)}]{haldane_hierarchy}%
  \BibitemOpen
  \bibfield  {author} {\bibinfo {author} {\bibfnamefont {F.~D.~M.}\
  \bibnamefont {Haldane}},\ }\href {\doibase 10.1103/PhysRevLett.51.605}
  {\bibfield  {journal} {\bibinfo  {journal} {Phys. Rev. Lett.}\ }\textbf
  {\bibinfo {volume} {51}},\ \bibinfo {pages} {605} (\bibinfo {year}
  {1983})}\BibitemShut {NoStop}%
\bibitem [{\citenamefont {Trugman}\ and\ \citenamefont {Kivelson}(1985)}]{TK}%
  \BibitemOpen
  \bibfield  {author} {\bibinfo {author} {\bibfnamefont {S.~A.}\ \bibnamefont
  {Trugman}}\ and\ \bibinfo {author} {\bibfnamefont {S.}~\bibnamefont
  {Kivelson}},\ }\href {\doibase 10.1103/PhysRevB.31.5280} {\bibfield
  {journal} {\bibinfo  {journal} {Phys. Rev. B}\ }\textbf {\bibinfo {volume}
  {31}},\ \bibinfo {pages} {5280} (\bibinfo {year} {1985})}\BibitemShut
  {NoStop}%
\bibitem [{\citenamefont {Laughlin}(1983)}]{Laughlin}%
  \BibitemOpen
  \bibfield  {author} {\bibinfo {author} {\bibfnamefont {R.~B.}\ \bibnamefont
  {Laughlin}},\ }\href {\doibase 10.1103/PhysRevLett.50.1395} {\bibfield
  {journal} {\bibinfo  {journal} {Phys. Rev. Lett.}\ }\textbf {\bibinfo
  {volume} {50}},\ \bibinfo {pages} {1395} (\bibinfo {year}
  {1983})}\BibitemShut {NoStop}%
\bibitem [{\citenamefont {Moore}\ and\ \citenamefont {Read}(1991)}]{MR}%
  \BibitemOpen
  \bibfield  {author} {\bibinfo {author} {\bibfnamefont {G.}~\bibnamefont
  {Moore}}\ and\ \bibinfo {author} {\bibfnamefont {N.}~\bibnamefont {Read}},\
  }\href {\doibase http://dx.doi.org/10.1016/0550-3213(91)90407-O} {\bibfield
  {journal} {\bibinfo  {journal} {Nuclear Physics B}\ }\textbf {\bibinfo
  {volume} {360}},\ \bibinfo {pages} {362 } (\bibinfo {year}
  {1991})}\BibitemShut {NoStop}%
\bibitem [{\citenamefont {Read}\ and\ \citenamefont {Rezayi}(1999)}]{RR}%
  \BibitemOpen
  \bibfield  {author} {\bibinfo {author} {\bibfnamefont {N.}~\bibnamefont
  {Read}}\ and\ \bibinfo {author} {\bibfnamefont {E.}~\bibnamefont {Rezayi}},\
  }\href {\doibase 10.1103/PhysRevB.59.8084} {\bibfield  {journal} {\bibinfo
  {journal} {Phys. Rev. B}\ }\textbf {\bibinfo {volume} {59}},\ \bibinfo
  {pages} {8084} (\bibinfo {year} {1999})}\BibitemShut {NoStop}%
\bibitem [{\citenamefont {Bernevig}\ and\ \citenamefont
  {Haldane}(2008{\natexlab{a}})}]{BH1}%
  \BibitemOpen
  \bibfield  {author} {\bibinfo {author} {\bibfnamefont {B.~A.}\ \bibnamefont
  {Bernevig}}\ and\ \bibinfo {author} {\bibfnamefont {F.~D.~M.}\ \bibnamefont
  {Haldane}},\ }\href {\doibase 10.1103/PhysRevB.77.184502} {\bibfield
  {journal} {\bibinfo  {journal} {Phys. Rev. B}\ }\textbf {\bibinfo {volume}
  {77}},\ \bibinfo {pages} {184502} (\bibinfo {year}
  {2008}{\natexlab{a}})}\BibitemShut {NoStop}%
\bibitem [{\citenamefont {Bernevig}\ and\ \citenamefont
  {Haldane}(2008{\natexlab{b}})}]{BH2}%
  \BibitemOpen
  \bibfield  {author} {\bibinfo {author} {\bibfnamefont {B.~A.}\ \bibnamefont
  {Bernevig}}\ and\ \bibinfo {author} {\bibfnamefont {F.~D.~M.}\ \bibnamefont
  {Haldane}},\ }\href {\doibase 10.1103/PhysRevLett.100.246802} {\bibfield
  {journal} {\bibinfo  {journal} {Phys. Rev. Lett.}\ }\textbf {\bibinfo
  {volume} {100}},\ \bibinfo {pages} {246802} (\bibinfo {year}
  {2008}{\natexlab{b}})}\BibitemShut {NoStop}%
\bibitem [{\citenamefont {Bernevig}\ and\ \citenamefont {Haldane}(2009)}]{BH3}%
  \BibitemOpen
  \bibfield  {author} {\bibinfo {author} {\bibfnamefont {B.~A.}\ \bibnamefont
  {Bernevig}}\ and\ \bibinfo {author} {\bibfnamefont {F.~D.~M.}\ \bibnamefont
  {Haldane}},\ }\href {\doibase 10.1103/PhysRevLett.102.066802} {\bibfield
  {journal} {\bibinfo  {journal} {Phys. Rev. Lett.}\ }\textbf {\bibinfo
  {volume} {102}},\ \bibinfo {pages} {066802} (\bibinfo {year}
  {2009})}\BibitemShut {NoStop}%
\bibitem [{\citenamefont {Wen}\ and\ \citenamefont
  {Wang}(2008{\natexlab{a}})}]{WW1}%
  \BibitemOpen
  \bibfield  {author} {\bibinfo {author} {\bibfnamefont {X.-G.}\ \bibnamefont
  {Wen}}\ and\ \bibinfo {author} {\bibfnamefont {Z.}~\bibnamefont {Wang}},\
  }\href {\doibase 10.1103/PhysRevB.77.235108} {\bibfield  {journal} {\bibinfo
  {journal} {Phys. Rev. B}\ }\textbf {\bibinfo {volume} {77}},\ \bibinfo
  {pages} {235108} (\bibinfo {year} {2008}{\natexlab{a}})}\BibitemShut
  {NoStop}%
\bibitem [{\citenamefont {Wen}\ and\ \citenamefont
  {Wang}(2008{\natexlab{b}})}]{WW2}%
  \BibitemOpen
  \bibfield  {author} {\bibinfo {author} {\bibfnamefont {X.-G.}\ \bibnamefont
  {Wen}}\ and\ \bibinfo {author} {\bibfnamefont {Z.}~\bibnamefont {Wang}},\
  }\href {\doibase 10.1103/PhysRevB.78.155109} {\bibfield  {journal} {\bibinfo
  {journal} {Phys. Rev. B}\ }\textbf {\bibinfo {volume} {78}},\ \bibinfo
  {pages} {155109} (\bibinfo {year} {2008}{\natexlab{b}})}\BibitemShut
  {NoStop}%
\bibitem [{\citenamefont {Simon}\ \emph {et~al.}(2007)\citenamefont {Simon},
  \citenamefont {Rezayi}, \citenamefont {Cooper},\ and\ \citenamefont
  {Berdnikov}}]{gaffnian}%
  \BibitemOpen
  \bibfield  {author} {\bibinfo {author} {\bibfnamefont {S.~H.}\ \bibnamefont
  {Simon}}, \bibinfo {author} {\bibfnamefont {E.~H.}\ \bibnamefont {Rezayi}},
  \bibinfo {author} {\bibfnamefont {N.~R.}\ \bibnamefont {Cooper}}, \ and\
  \bibinfo {author} {\bibfnamefont {I.}~\bibnamefont {Berdnikov}},\ }\href
  {\doibase 10.1103/PhysRevB.75.075317} {\bibfield  {journal} {\bibinfo
  {journal} {Phys. Rev. B}\ }\textbf {\bibinfo {volume} {75}},\ \bibinfo
  {pages} {075317} (\bibinfo {year} {2007})}\BibitemShut {NoStop}%
\bibitem [{\citenamefont {Greiter}\ \emph {et~al.}(2014)\citenamefont
  {Greiter}, \citenamefont {Schroeter},\ and\ \citenamefont
  {Thomale}}]{greiter}%
  \BibitemOpen
  \bibfield  {author} {\bibinfo {author} {\bibfnamefont {M.}~\bibnamefont
  {Greiter}}, \bibinfo {author} {\bibfnamefont {D.~F.}\ \bibnamefont
  {Schroeter}}, \ and\ \bibinfo {author} {\bibfnamefont {R.}~\bibnamefont
  {Thomale}},\ }\href {\doibase 10.1103/PhysRevB.89.165125} {\bibfield
  {journal} {\bibinfo  {journal} {Phys. Rev. B}\ }\textbf {\bibinfo {volume}
  {89}},\ \bibinfo {pages} {165125} (\bibinfo {year} {2014})}\BibitemShut
  {NoStop}%
\bibitem [{\citenamefont {Read}(2006)}]{RRcount}%
  \BibitemOpen
  \bibfield  {author} {\bibinfo {author} {\bibfnamefont {N.}~\bibnamefont
  {Read}},\ }\href {\doibase 10.1103/PhysRevB.73.245334} {\bibfield  {journal}
  {\bibinfo  {journal} {Phys. Rev. B}\ }\textbf {\bibinfo {volume} {73}},\
  \bibinfo {pages} {245334} (\bibinfo {year} {2006})}\BibitemShut {NoStop}%
\bibitem [{\citenamefont {Milovanovi\ifmmode~\acute{c}\else \'{c}\fi{}}\ and\
  \citenamefont {Read}(1996)}]{milovanovic}%
  \BibitemOpen
  \bibfield  {author} {\bibinfo {author} {\bibfnamefont {M.}~\bibnamefont
  {Milovanovi\ifmmode~\acute{c}\else \'{c}\fi{}}}\ and\ \bibinfo {author}
  {\bibfnamefont {N.}~\bibnamefont {Read}},\ }\href {\doibase
  10.1103/PhysRevB.53.13559} {\bibfield  {journal} {\bibinfo  {journal} {Phys.
  Rev. B}\ }\textbf {\bibinfo {volume} {53}},\ \bibinfo {pages} {13559}
  (\bibinfo {year} {1996})}\BibitemShut {NoStop}%
\bibitem [{\citenamefont {Ardonne}\ \emph {et~al.}(2001)\citenamefont
  {Ardonne}, \citenamefont {Read}, \citenamefont {Rezayi},\ and\ \citenamefont
  {Schoutens}}]{ardonne}%
  \BibitemOpen
  \bibfield  {author} {\bibinfo {author} {\bibfnamefont {E.}~\bibnamefont
  {Ardonne}}, \bibinfo {author} {\bibfnamefont {N.}~\bibnamefont {Read}},
  \bibinfo {author} {\bibfnamefont {E.}~\bibnamefont {Rezayi}}, \ and\ \bibinfo
  {author} {\bibfnamefont {K.}~\bibnamefont {Schoutens}},\ }\href {\doibase
  http://dx.doi.org/10.1016/S0550-3213(01)00224-3} {\bibfield  {journal}
  {\bibinfo  {journal} {Nuclear Physics B}\ }\textbf {\bibinfo {volume}
  {607}},\ \bibinfo {pages} {549 } (\bibinfo {year} {2001})}\BibitemShut
  {NoStop}%
\bibitem [{\citenamefont {Chandran}\ \emph {et~al.}(2011)\citenamefont
  {Chandran}, \citenamefont {Hermanns}, \citenamefont {Regnault},\ and\
  \citenamefont {Bernevig}}]{CHRB}%
  \BibitemOpen
  \bibfield  {author} {\bibinfo {author} {\bibfnamefont {A.}~\bibnamefont
  {Chandran}}, \bibinfo {author} {\bibfnamefont {M.}~\bibnamefont {Hermanns}},
  \bibinfo {author} {\bibfnamefont {N.}~\bibnamefont {Regnault}}, \ and\
  \bibinfo {author} {\bibfnamefont {B.~A.}\ \bibnamefont {Bernevig}},\ }\href
  {\doibase 10.1103/PhysRevB.84.205136} {\bibfield  {journal} {\bibinfo
  {journal} {Phys. Rev. B}\ }\textbf {\bibinfo {volume} {84}},\ \bibinfo
  {pages} {205136} (\bibinfo {year} {2011})}\BibitemShut {NoStop}%
\bibitem [{\citenamefont {Ortiz}\ \emph {et~al.}(2013)\citenamefont {Ortiz},
  \citenamefont {Nussinov}, \citenamefont {Dukelsky},\ and\ \citenamefont
  {Seidel}}]{ortiz}%
  \BibitemOpen
  \bibfield  {author} {\bibinfo {author} {\bibfnamefont {G.}~\bibnamefont
  {Ortiz}}, \bibinfo {author} {\bibfnamefont {Z.}~\bibnamefont {Nussinov}},
  \bibinfo {author} {\bibfnamefont {J.}~\bibnamefont {Dukelsky}}, \ and\
  \bibinfo {author} {\bibfnamefont {A.}~\bibnamefont {Seidel}},\ }\href
  {\doibase 10.1103/PhysRevB.88.165303} {\bibfield  {journal} {\bibinfo
  {journal} {Phys. Rev. B}\ }\textbf {\bibinfo {volume} {88}},\ \bibinfo
  {pages} {165303} (\bibinfo {year} {2013})}\BibitemShut {NoStop}%
\bibitem [{\citenamefont {Lee}\ and\ \citenamefont {Leinaas}(2004)}]{LL}%
  \BibitemOpen
  \bibfield  {author} {\bibinfo {author} {\bibfnamefont {D.-H.}\ \bibnamefont
  {Lee}}\ and\ \bibinfo {author} {\bibfnamefont {J.~M.}\ \bibnamefont
  {Leinaas}},\ }\href {\doibase 10.1103/PhysRevLett.92.096401} {\bibfield
  {journal} {\bibinfo  {journal} {Phys. Rev. Lett.}\ }\textbf {\bibinfo
  {volume} {92}},\ \bibinfo {pages} {096401} (\bibinfo {year}
  {2004})}\BibitemShut {NoStop}%
\bibitem [{\citenamefont {Seidel}\ \emph {et~al.}(2005)\citenamefont {Seidel},
  \citenamefont {Fu}, \citenamefont {Lee}, \citenamefont {Leinaas},\ and\
  \citenamefont {Moore}}]{seidel05}%
  \BibitemOpen
  \bibfield  {author} {\bibinfo {author} {\bibfnamefont {A.}~\bibnamefont
  {Seidel}}, \bibinfo {author} {\bibfnamefont {H.}~\bibnamefont {Fu}}, \bibinfo
  {author} {\bibfnamefont {D.-H.}\ \bibnamefont {Lee}}, \bibinfo {author}
  {\bibfnamefont {J.~M.}\ \bibnamefont {Leinaas}}, \ and\ \bibinfo {author}
  {\bibfnamefont {J.}~\bibnamefont {Moore}},\ }\href {\doibase
  10.1103/PhysRevLett.95.266405} {\bibfield  {journal} {\bibinfo  {journal}
  {Phys. Rev. Lett.}\ }\textbf {\bibinfo {volume} {95}},\ \bibinfo {pages}
  {266405} (\bibinfo {year} {2005})}\BibitemShut {NoStop}%
\bibitem [{\citenamefont {Qi}(2011)}]{qi}%
  \BibitemOpen
  \bibfield  {author} {\bibinfo {author} {\bibfnamefont {X.-L.}\ \bibnamefont
  {Qi}},\ }\href {\doibase 10.1103/PhysRevLett.107.126803} {\bibfield
  {journal} {\bibinfo  {journal} {Phys. Rev. Lett.}\ }\textbf {\bibinfo
  {volume} {107}},\ \bibinfo {pages} {126803} (\bibinfo {year}
  {2011})}\BibitemShut {NoStop}%
\bibitem [{\citenamefont {Tang}\ \emph {et~al.}(2011)\citenamefont {Tang},
  \citenamefont {Mei},\ and\ \citenamefont {Wen}}]{Tang11}%
  \BibitemOpen
  \bibfield  {author} {\bibinfo {author} {\bibfnamefont {E.}~\bibnamefont
  {Tang}}, \bibinfo {author} {\bibfnamefont {J.-W.}\ \bibnamefont {Mei}}, \
  and\ \bibinfo {author} {\bibfnamefont {X.-G.}\ \bibnamefont {Wen}},\ }\href
  {\doibase 10.1103/PhysRevLett.106.236802} {\bibfield  {journal} {\bibinfo
  {journal} {Phys. Rev. Lett.}\ }\textbf {\bibinfo {volume} {106}},\ \bibinfo
  {pages} {236802} (\bibinfo {year} {2011})}\BibitemShut {NoStop}%
\bibitem [{\citenamefont {Sun}\ \emph {et~al.}(2011)\citenamefont {Sun},
  \citenamefont {Gu}, \citenamefont {Katsura},\ and\ \citenamefont
  {Das~Sarma}}]{Sun11}%
  \BibitemOpen
  \bibfield  {author} {\bibinfo {author} {\bibfnamefont {K.}~\bibnamefont
  {Sun}}, \bibinfo {author} {\bibfnamefont {Z.}~\bibnamefont {Gu}}, \bibinfo
  {author} {\bibfnamefont {H.}~\bibnamefont {Katsura}}, \ and\ \bibinfo
  {author} {\bibfnamefont {S.}~\bibnamefont {Das~Sarma}},\ }\href {\doibase
  10.1103/PhysRevLett.106.236803} {\bibfield  {journal} {\bibinfo  {journal}
  {Phys. Rev. Lett.}\ }\textbf {\bibinfo {volume} {106}},\ \bibinfo {pages}
  {236803} (\bibinfo {year} {2011})}\BibitemShut {NoStop}%
\bibitem [{\citenamefont {Neupert}\ \emph {et~al.}(2011)\citenamefont
  {Neupert}, \citenamefont {Santos}, \citenamefont {Chamon},\ and\
  \citenamefont {Mudry}}]{Neu11}%
  \BibitemOpen
  \bibfield  {author} {\bibinfo {author} {\bibfnamefont {T.}~\bibnamefont
  {Neupert}}, \bibinfo {author} {\bibfnamefont {L.}~\bibnamefont {Santos}},
  \bibinfo {author} {\bibfnamefont {C.}~\bibnamefont {Chamon}}, \ and\ \bibinfo
  {author} {\bibfnamefont {C.}~\bibnamefont {Mudry}},\ }\href {\doibase
  10.1103/PhysRevLett.106.236804} {\bibfield  {journal} {\bibinfo  {journal}
  {Phys. Rev. Lett.}\ }\textbf {\bibinfo {volume} {106}},\ \bibinfo {pages}
  {236804} (\bibinfo {year} {2011})}\BibitemShut {NoStop}%
\bibitem [{\citenamefont {{Sheng}}\ \emph {et~al.}(2011)\citenamefont
  {{Sheng}}, \citenamefont {{Gu}}, \citenamefont {{Sun}},\ and\ \citenamefont
  {{Sheng}}}]{Sheng11}%
  \BibitemOpen
  \bibfield  {author} {\bibinfo {author} {\bibfnamefont {D.~N.}\ \bibnamefont
  {{Sheng}}}, \bibinfo {author} {\bibfnamefont {Z.-C.}\ \bibnamefont {{Gu}}},
  \bibinfo {author} {\bibfnamefont {K.}~\bibnamefont {{Sun}}}, \ and\ \bibinfo
  {author} {\bibfnamefont {L.}~\bibnamefont {{Sheng}}},\ }\href {\doibase
  10.1038/ncomms1380} {\bibfield  {journal} {\bibinfo  {journal} {Nature
  Communications}\ }\textbf {\bibinfo {volume} {2}},\ \bibinfo {eid} {389}
  (\bibinfo {year} {2011}),\ 10.1038/ncomms1380},\ \Eprint
  {http://arxiv.org/abs/1102.2658} {arXiv:1102.2658 [cond-mat.str-el]}
  \BibitemShut {NoStop}%
\bibitem [{\citenamefont {Wang}\ \emph {et~al.}(2011)\citenamefont {Wang},
  \citenamefont {Gu}, \citenamefont {Gong},\ and\ \citenamefont
  {Sheng}}]{Wang11}%
  \BibitemOpen
  \bibfield  {author} {\bibinfo {author} {\bibfnamefont {Y.-F.}\ \bibnamefont
  {Wang}}, \bibinfo {author} {\bibfnamefont {Z.-C.}\ \bibnamefont {Gu}},
  \bibinfo {author} {\bibfnamefont {C.-D.}\ \bibnamefont {Gong}}, \ and\
  \bibinfo {author} {\bibfnamefont {D.~N.}\ \bibnamefont {Sheng}},\ }\href
  {\doibase 10.1103/PhysRevLett.107.146803} {\bibfield  {journal} {\bibinfo
  {journal} {Phys. Rev. Lett.}\ }\textbf {\bibinfo {volume} {107}},\ \bibinfo
  {pages} {146803} (\bibinfo {year} {2011})}\BibitemShut {NoStop}%
\bibitem [{\citenamefont {Regnault}\ and\ \citenamefont
  {Bernevig}(2011)}]{Regnault11}%
  \BibitemOpen
  \bibfield  {author} {\bibinfo {author} {\bibfnamefont {N.}~\bibnamefont
  {Regnault}}\ and\ \bibinfo {author} {\bibfnamefont {B.~A.}\ \bibnamefont
  {Bernevig}},\ }\href {\doibase 10.1103/PhysRevX.1.021014} {\bibfield
  {journal} {\bibinfo  {journal} {Phys. Rev. X}\ }\textbf {\bibinfo {volume}
  {1}},\ \bibinfo {pages} {021014} (\bibinfo {year} {2011})}\BibitemShut
  {NoStop}%
\bibitem [{\citenamefont {Wang}\ and\ \citenamefont {Ran}(2011)}]{Ran11}%
  \BibitemOpen
  \bibfield  {author} {\bibinfo {author} {\bibfnamefont {F.}~\bibnamefont
  {Wang}}\ and\ \bibinfo {author} {\bibfnamefont {Y.}~\bibnamefont {Ran}},\
  }\href {\doibase 10.1103/PhysRevB.84.241103} {\bibfield  {journal} {\bibinfo
  {journal} {Phys. Rev. B}\ }\textbf {\bibinfo {volume} {84}},\ \bibinfo
  {pages} {241103} (\bibinfo {year} {2011})}\BibitemShut {NoStop}%
\bibitem [{\citenamefont {Bernevig}\ and\ \citenamefont
  {Regnault}(2012)}]{Bernevig12}%
  \BibitemOpen
  \bibfield  {author} {\bibinfo {author} {\bibfnamefont {B.~A.}\ \bibnamefont
  {Bernevig}}\ and\ \bibinfo {author} {\bibfnamefont {N.}~\bibnamefont
  {Regnault}},\ }\href {\doibase 10.1103/PhysRevB.85.075128} {\bibfield
  {journal} {\bibinfo  {journal} {Phys. Rev. B}\ }\textbf {\bibinfo {volume}
  {85}},\ \bibinfo {pages} {075128} (\bibinfo {year} {2012})}\BibitemShut
  {NoStop}%
\bibitem [{\citenamefont {Wang}\ \emph
  {et~al.}(2012{\natexlab{a}})\citenamefont {Wang}, \citenamefont {Yao},
  \citenamefont {Gu}, \citenamefont {Gong},\ and\ \citenamefont
  {Sheng}}]{WangPRL12}%
  \BibitemOpen
  \bibfield  {author} {\bibinfo {author} {\bibfnamefont {Y.-F.}\ \bibnamefont
  {Wang}}, \bibinfo {author} {\bibfnamefont {H.}~\bibnamefont {Yao}}, \bibinfo
  {author} {\bibfnamefont {Z.-C.}\ \bibnamefont {Gu}}, \bibinfo {author}
  {\bibfnamefont {C.-D.}\ \bibnamefont {Gong}}, \ and\ \bibinfo {author}
  {\bibfnamefont {D.~N.}\ \bibnamefont {Sheng}},\ }\href {\doibase
  10.1103/PhysRevLett.108.126805} {\bibfield  {journal} {\bibinfo  {journal}
  {Phys. Rev. Lett.}\ }\textbf {\bibinfo {volume} {108}},\ \bibinfo {pages}
  {126805} (\bibinfo {year} {2012}{\natexlab{a}})}\BibitemShut {NoStop}%
\bibitem [{\citenamefont {Wang}\ \emph
  {et~al.}(2012{\natexlab{b}})\citenamefont {Wang}, \citenamefont {Yao},
  \citenamefont {Gong},\ and\ \citenamefont {Sheng}}]{WangPRB12}%
  \BibitemOpen
  \bibfield  {author} {\bibinfo {author} {\bibfnamefont {Y.-F.}\ \bibnamefont
  {Wang}}, \bibinfo {author} {\bibfnamefont {H.}~\bibnamefont {Yao}}, \bibinfo
  {author} {\bibfnamefont {C.-D.}\ \bibnamefont {Gong}}, \ and\ \bibinfo
  {author} {\bibfnamefont {D.~N.}\ \bibnamefont {Sheng}},\ }\href {\doibase
  10.1103/PhysRevB.86.201101} {\bibfield  {journal} {\bibinfo  {journal} {Phys.
  Rev. B}\ }\textbf {\bibinfo {volume} {86}},\ \bibinfo {pages} {201101}
  (\bibinfo {year} {2012}{\natexlab{b}})}\BibitemShut {NoStop}%
\bibitem [{\citenamefont {Wu}\ \emph {et~al.}(2012)\citenamefont {Wu},
  \citenamefont {Bernevig},\ and\ \citenamefont {Regnault}}]{Wu12}%
  \BibitemOpen
  \bibfield  {author} {\bibinfo {author} {\bibfnamefont {Y.-L.}\ \bibnamefont
  {Wu}}, \bibinfo {author} {\bibfnamefont {B.~A.}\ \bibnamefont {Bernevig}}, \
  and\ \bibinfo {author} {\bibfnamefont {N.}~\bibnamefont {Regnault}},\ }\href
  {\doibase 10.1103/PhysRevB.85.075116} {\bibfield  {journal} {\bibinfo
  {journal} {Phys. Rev. B}\ }\textbf {\bibinfo {volume} {85}},\ \bibinfo
  {pages} {075116} (\bibinfo {year} {2012})}\BibitemShut {NoStop}%
\bibitem [{\citenamefont {Liu}\ \emph {et~al.}(2012)\citenamefont {Liu},
  \citenamefont {Bergholtz}, \citenamefont {Fan},\ and\ \citenamefont
  {L\"auchli}}]{Laeuchli12}%
  \BibitemOpen
  \bibfield  {author} {\bibinfo {author} {\bibfnamefont {Z.}~\bibnamefont
  {Liu}}, \bibinfo {author} {\bibfnamefont {E.~J.}\ \bibnamefont {Bergholtz}},
  \bibinfo {author} {\bibfnamefont {H.}~\bibnamefont {Fan}}, \ and\ \bibinfo
  {author} {\bibfnamefont {A.~M.}\ \bibnamefont {L\"auchli}},\ }\href {\doibase
  10.1103/PhysRevLett.109.186805} {\bibfield  {journal} {\bibinfo  {journal}
  {Phys. Rev. Lett.}\ }\textbf {\bibinfo {volume} {109}},\ \bibinfo {pages}
  {186805} (\bibinfo {year} {2012})}\BibitemShut {NoStop}%
\bibitem [{\citenamefont {Trescher}\ and\ \citenamefont
  {Bergholtz}(2012)}]{Bergholtz12}%
  \BibitemOpen
  \bibfield  {author} {\bibinfo {author} {\bibfnamefont {M.}~\bibnamefont
  {Trescher}}\ and\ \bibinfo {author} {\bibfnamefont {E.~J.}\ \bibnamefont
  {Bergholtz}},\ }\href {\doibase 10.1103/PhysRevB.86.241111} {\bibfield
  {journal} {\bibinfo  {journal} {Phys. Rev. B}\ }\textbf {\bibinfo {volume}
  {86}},\ \bibinfo {pages} {241111} (\bibinfo {year} {2012})}\BibitemShut
  {NoStop}%
\bibitem [{\citenamefont {Yang}\ \emph {et~al.}(2012)\citenamefont {Yang},
  \citenamefont {Gu}, \citenamefont {Sun},\ and\ \citenamefont
  {Das~Sarma}}]{Sarma12}%
  \BibitemOpen
  \bibfield  {author} {\bibinfo {author} {\bibfnamefont {S.}~\bibnamefont
  {Yang}}, \bibinfo {author} {\bibfnamefont {Z.-C.}\ \bibnamefont {Gu}},
  \bibinfo {author} {\bibfnamefont {K.}~\bibnamefont {Sun}}, \ and\ \bibinfo
  {author} {\bibfnamefont {S.}~\bibnamefont {Das~Sarma}},\ }\href {\doibase
  10.1103/PhysRevB.86.241112} {\bibfield  {journal} {\bibinfo  {journal} {Phys.
  Rev. B}\ }\textbf {\bibinfo {volume} {86}},\ \bibinfo {pages} {241112}
  (\bibinfo {year} {2012})}\BibitemShut {NoStop}%
\bibitem [{\citenamefont {Liu}\ \emph {et~al.}(2013)\citenamefont {Liu},
  \citenamefont {Repellin}, \citenamefont {Bernevig},\ and\ \citenamefont
  {Regnault}}]{Liu13}%
  \BibitemOpen
  \bibfield  {author} {\bibinfo {author} {\bibfnamefont {T.}~\bibnamefont
  {Liu}}, \bibinfo {author} {\bibfnamefont {C.}~\bibnamefont {Repellin}},
  \bibinfo {author} {\bibfnamefont {B.~A.}\ \bibnamefont {Bernevig}}, \ and\
  \bibinfo {author} {\bibfnamefont {N.}~\bibnamefont {Regnault}},\ }\href
  {\doibase 10.1103/PhysRevB.87.205136} {\bibfield  {journal} {\bibinfo
  {journal} {Phys. Rev. B}\ }\textbf {\bibinfo {volume} {87}},\ \bibinfo
  {pages} {205136} (\bibinfo {year} {2013})}\BibitemShut {NoStop}%
\bibitem [{\citenamefont {Chen}\ \emph {et~al.}(2014)\citenamefont {Chen},
  \citenamefont {Mazaheri}, \citenamefont {Seidel},\ and\ \citenamefont
  {Tang}}]{chen13}%
  \BibitemOpen
  \bibfield  {author} {\bibinfo {author} {\bibfnamefont {L.}~\bibnamefont
  {Chen}}, \bibinfo {author} {\bibfnamefont {T.}~\bibnamefont {Mazaheri}},
  \bibinfo {author} {\bibfnamefont {A.}~\bibnamefont {Seidel}}, \ and\ \bibinfo
  {author} {\bibfnamefont {X.}~\bibnamefont {Tang}},\ }\href
  {http://stacks.iop.org/1751-8121/47/i=15/a=152001} {\bibfield  {journal}
  {\bibinfo  {journal} {Journal of Physics A: Mathematical and Theoretical}\
  }\textbf {\bibinfo {volume} {47}},\ \bibinfo {pages} {152001} (\bibinfo
  {year} {2014})}\BibitemShut {NoStop}%
\bibitem [{\citenamefont {Scaffidi}\ and\ \citenamefont
  {Simon}(2014)}]{scaffidi}%
  \BibitemOpen
  \bibfield  {author} {\bibinfo {author} {\bibfnamefont {T.}~\bibnamefont
  {Scaffidi}}\ and\ \bibinfo {author} {\bibfnamefont {S.~H.}\ \bibnamefont
  {Simon}},\ }\href {\doibase 10.1103/PhysRevB.90.115132} {\bibfield  {journal}
  {\bibinfo  {journal} {Phys. Rev. B}\ }\textbf {\bibinfo {volume} {90}},\
  \bibinfo {pages} {115132} (\bibinfo {year} {2014})}\BibitemShut {NoStop}%
\bibitem [{\citenamefont {Lee}\ \emph {et~al.}(2013)\citenamefont {Lee},
  \citenamefont {Thomale},\ and\ \citenamefont {Qi}}]{CHLee}%
  \BibitemOpen
  \bibfield  {author} {\bibinfo {author} {\bibfnamefont {C.~H.}\ \bibnamefont
  {Lee}}, \bibinfo {author} {\bibfnamefont {R.}~\bibnamefont {Thomale}}, \ and\
  \bibinfo {author} {\bibfnamefont {X.-L.}\ \bibnamefont {Qi}},\ }\href
  {\doibase 10.1103/PhysRevB.88.035101} {\bibfield  {journal} {\bibinfo
  {journal} {Phys. Rev. B}\ }\textbf {\bibinfo {volume} {88}},\ \bibinfo
  {pages} {035101} (\bibinfo {year} {2013})}\BibitemShut {NoStop}%
\bibitem [{\citenamefont {Haldane}(2011)}]{Haldane11}%
  \BibitemOpen
  \bibfield  {author} {\bibinfo {author} {\bibfnamefont {F.~D.~M.}\
  \bibnamefont {Haldane}},\ }\href {\doibase 10.1103/PhysRevLett.107.116801}
  {\bibfield  {journal} {\bibinfo  {journal} {Phys. Rev. Lett.}\ }\textbf
  {\bibinfo {volume} {107}},\ \bibinfo {pages} {116801} (\bibinfo {year}
  {2011})}\BibitemShut {NoStop}%
\bibitem [{\citenamefont {Affleck}\ \emph {et~al.}(1988)\citenamefont
  {Affleck}, \citenamefont {Kennedy}, \citenamefont {Lieb},\ and\ \citenamefont
  {Tasaki}}]{AKLT}%
  \BibitemOpen
  \bibfield  {author} {\bibinfo {author} {\bibfnamefont {I.}~\bibnamefont
  {Affleck}}, \bibinfo {author} {\bibfnamefont {T.}~\bibnamefont {Kennedy}},
  \bibinfo {author} {\bibfnamefont {E.}~\bibnamefont {Lieb}}, \ and\ \bibinfo
  {author} {\bibfnamefont {H.}~\bibnamefont {Tasaki}},\ }\href {\doibase
  10.1007/BF01218021} {\bibfield  {journal} {\bibinfo  {journal}
  {Communications in Mathematical Physics}\ }\textbf {\bibinfo {volume}
  {115}},\ \bibinfo {pages} {477} (\bibinfo {year} {1988})}\BibitemShut
  {NoStop}%
\bibitem [{\citenamefont {Fannes}\ \emph {et~al.}(1992)\citenamefont {Fannes},
  \citenamefont {Nachtergaele},\ and\ \citenamefont {Werner}}]{nachtergaele}%
  \BibitemOpen
  \bibfield  {author} {\bibinfo {author} {\bibfnamefont {M.}~\bibnamefont
  {Fannes}}, \bibinfo {author} {\bibfnamefont {B.}~\bibnamefont
  {Nachtergaele}}, \ and\ \bibinfo {author} {\bibfnamefont {R.}~\bibnamefont
  {Werner}},\ }\href {\doibase 10.1007/BF02099178} {\bibfield  {journal}
  {\bibinfo  {journal} {Communications in Mathematical Physics}\ }\textbf
  {\bibinfo {volume} {144}},\ \bibinfo {pages} {443} (\bibinfo {year}
  {1992})}\BibitemShut {NoStop}%
\bibitem [{\citenamefont {Kennedy}\ and\ \citenamefont
  {Tasaki}(1992)}]{Kennedy92}%
  \BibitemOpen
  \bibfield  {author} {\bibinfo {author} {\bibfnamefont {T.}~\bibnamefont
  {Kennedy}}\ and\ \bibinfo {author} {\bibfnamefont {H.}~\bibnamefont
  {Tasaki}},\ }\href {\doibase 10.1007/BF02097239} {\bibfield  {journal}
  {\bibinfo  {journal} {Communications in Mathematical Physics}\ }\textbf
  {\bibinfo {volume} {147}},\ \bibinfo {pages} {431} (\bibinfo {year}
  {1992})}\BibitemShut {NoStop}%
\bibitem [{\citenamefont {Bravyi}\ and\ \citenamefont
  {Terhal}(2010)}]{Bravyi09}%
  \BibitemOpen
  \bibfield  {author} {\bibinfo {author} {\bibfnamefont {S.}~\bibnamefont
  {Bravyi}}\ and\ \bibinfo {author} {\bibfnamefont {B.}~\bibnamefont
  {Terhal}},\ }\href {\doibase 10.1137/08072689X} {\bibfield  {journal}
  {\bibinfo  {journal} {SIAM Journal on Computing}\ }\textbf {\bibinfo {volume}
  {39}},\ \bibinfo {pages} {1462} (\bibinfo {year} {2010})},\ \Eprint
  {http://arxiv.org/abs/http://dx.doi.org/10.1137/08072689X}
  {http://dx.doi.org/10.1137/08072689X} \BibitemShut {NoStop}%
\bibitem [{\citenamefont {de~Beaudrap}\ \emph {et~al.}(2010)\citenamefont
  {de~Beaudrap}, \citenamefont {Osborne},\ and\ \citenamefont {Eisert}}]{NdB}%
  \BibitemOpen
  \bibfield  {author} {\bibinfo {author} {\bibfnamefont {N.}~\bibnamefont
  {de~Beaudrap}}, \bibinfo {author} {\bibfnamefont {T.~J.}\ \bibnamefont
  {Osborne}}, \ and\ \bibinfo {author} {\bibfnamefont {J.}~\bibnamefont
  {Eisert}},\ }\href {http://stacks.iop.org/1367-2630/12/i=9/a=095007}
  {\bibfield  {journal} {\bibinfo  {journal} {New Journal of Physics}\ }\textbf
  {\bibinfo {volume} {12}},\ \bibinfo {pages} {095007} (\bibinfo {year}
  {2010})}\BibitemShut {NoStop}%
\bibitem [{\citenamefont {Yoshida}(2011)}]{Yoshida}%
  \BibitemOpen
  \bibfield  {author} {\bibinfo {author} {\bibfnamefont {B.}~\bibnamefont
  {Yoshida}},\ }\href {\doibase http://dx.doi.org/10.1016/j.aop.2010.10.009}
  {\bibfield  {journal} {\bibinfo  {journal} {Annals of Physics}\ }\textbf
  {\bibinfo {volume} {326}},\ \bibinfo {pages} {15 } (\bibinfo {year}
  {2011})},\ \bibinfo {note} {january 2011 Special Issue}\BibitemShut {NoStop}%
\bibitem [{\citenamefont {Chen}\ \emph {et~al.}(2012)\citenamefont {Chen},
  \citenamefont {Ji}, \citenamefont {Kribs}, \citenamefont {Wei},\ and\
  \citenamefont {Zeng}}]{CJKWZ}%
  \BibitemOpen
  \bibfield  {author} {\bibinfo {author} {\bibfnamefont {J.}~\bibnamefont
  {Chen}}, \bibinfo {author} {\bibfnamefont {Z.}~\bibnamefont {Ji}}, \bibinfo
  {author} {\bibfnamefont {D.}~\bibnamefont {Kribs}}, \bibinfo {author}
  {\bibfnamefont {Z.}~\bibnamefont {Wei}}, \ and\ \bibinfo {author}
  {\bibfnamefont {B.}~\bibnamefont {Zeng}},\ }\href {\doibase
  http://dx.doi.org/10.1063/1.4748527} {\bibfield  {journal} {\bibinfo
  {journal} {Journal of Mathematical Physics}\ }\textbf {\bibinfo {volume}
  {53}},\ \bibinfo {eid} {102201} (\bibinfo {year} {2012})}\BibitemShut
  {NoStop}%
\bibitem [{\citenamefont {Michalakis}\ and\ \citenamefont {Zwolak}(2013)}]{MZ}%
  \BibitemOpen
  \bibfield  {author} {\bibinfo {author} {\bibfnamefont {S.}~\bibnamefont
  {Michalakis}}\ and\ \bibinfo {author} {\bibfnamefont {J.}~\bibnamefont
  {Zwolak}},\ }\href {\doibase 10.1007/s00220-013-1762-6} {\bibfield  {journal}
  {\bibinfo  {journal} {Communications in Mathematical Physics}\ }\textbf
  {\bibinfo {volume} {322}},\ \bibinfo {pages} {277} (\bibinfo {year}
  {2013})}\BibitemShut {NoStop}%
\bibitem [{\citenamefont {Schuch}\ \emph {et~al.}(2010)\citenamefont {Schuch},
  \citenamefont {Cirac},\ and\ \citenamefont {Perez-Garcia}}]{schuch}%
  \BibitemOpen
  \bibfield  {author} {\bibinfo {author} {\bibfnamefont {N.}~\bibnamefont
  {Schuch}}, \bibinfo {author} {\bibfnamefont {I.}~\bibnamefont {Cirac}}, \
  and\ \bibinfo {author} {\bibfnamefont {D.}~\bibnamefont {Perez-Garcia}},\
  }\href {\doibase http://dx.doi.org/10.1016/j.aop.2010.05.008} {\bibfield
  {journal} {\bibinfo  {journal} {Annals of Physics}\ }\textbf {\bibinfo
  {volume} {325}},\ \bibinfo {pages} {2153 } (\bibinfo {year}
  {2010})}\BibitemShut {NoStop}%
\bibitem [{\citenamefont {Bravyi}\ and\ \citenamefont
  {Hastings}(2011)}]{Bravyi}%
  \BibitemOpen
  \bibfield  {author} {\bibinfo {author} {\bibfnamefont {S.}~\bibnamefont
  {Bravyi}}\ and\ \bibinfo {author} {\bibfnamefont {M.}~\bibnamefont
  {Hastings}},\ }\href {\doibase 10.1007/s00220-011-1346-2} {\bibfield
  {journal} {\bibinfo  {journal} {Communications in Mathematical Physics}\
  }\textbf {\bibinfo {volume} {307}},\ \bibinfo {pages} {609} (\bibinfo {year}
  {2011})}\BibitemShut {NoStop}%
\bibitem [{\citenamefont {Darmawan}\ and\ \citenamefont
  {Bartlett}(2014)}]{Darmawan}%
  \BibitemOpen
  \bibfield  {author} {\bibinfo {author} {\bibfnamefont {A.~S.}\ \bibnamefont
  {Darmawan}}\ and\ \bibinfo {author} {\bibfnamefont {S.~D.}\ \bibnamefont
  {Bartlett}},\ }\href {http://stacks.iop.org/1367-2630/16/i=7/a=073013}
  {\bibfield  {journal} {\bibinfo  {journal} {New Journal of Physics}\ }\textbf
  {\bibinfo {volume} {16}},\ \bibinfo {pages} {073013} (\bibinfo {year}
  {2014})}\BibitemShut {NoStop}%
\bibitem [{\citenamefont {{Dubail}}\ \emph {et~al.}(2012)\citenamefont
  {{Dubail}}, \citenamefont {{Read}},\ and\ \citenamefont {{Rezayi}}}]{dubail}%
  \BibitemOpen
  \bibfield  {author} {\bibinfo {author} {\bibfnamefont {J.}~\bibnamefont
  {{Dubail}}}, \bibinfo {author} {\bibfnamefont {N.}~\bibnamefont {{Read}}}, \
  and\ \bibinfo {author} {\bibfnamefont {E.~H.}\ \bibnamefont {{Rezayi}}},\
  }\href {\doibase 10.1103/PhysRevB.86.245310} {\bibfield  {journal} {\bibinfo
  {journal} {\prb}\ }\textbf {\bibinfo {volume} {86}},\ \bibinfo {eid} {245310}
  (\bibinfo {year} {2012})},\ \Eprint {http://arxiv.org/abs/1207.7119}
  {arXiv:1207.7119 [cond-mat.mes-hall]} \BibitemShut {NoStop}%
\bibitem [{\citenamefont {{Zaletel}}\ and\ \citenamefont
  {{Mong}}(2012)}]{zaletel}%
  \BibitemOpen
  \bibfield  {author} {\bibinfo {author} {\bibfnamefont {M.~P.}\ \bibnamefont
  {{Zaletel}}}\ and\ \bibinfo {author} {\bibfnamefont {R.~S.~K.}\ \bibnamefont
  {{Mong}}},\ }\href {\doibase 10.1103/PhysRevB.86.245305} {\bibfield
  {journal} {\bibinfo  {journal} {\prb}\ }\textbf {\bibinfo {volume} {86}},\
  \bibinfo {eid} {245305} (\bibinfo {year} {2012})},\ \Eprint
  {http://arxiv.org/abs/1208.4862} {arXiv:1208.4862 [cond-mat.str-el]}
  \BibitemShut {NoStop}%
\bibitem [{\citenamefont {Read}\ and\ \citenamefont {Rezayi}(2011)}]{RR11}%
  \BibitemOpen
  \bibfield  {author} {\bibinfo {author} {\bibfnamefont {N.}~\bibnamefont
  {Read}}\ and\ \bibinfo {author} {\bibfnamefont {E.~H.}\ \bibnamefont
  {Rezayi}},\ }\href {\doibase 10.1103/PhysRevB.84.085316} {\bibfield
  {journal} {\bibinfo  {journal} {Phys. Rev. B}\ }\textbf {\bibinfo {volume}
  {84}},\ \bibinfo {pages} {085316} (\bibinfo {year} {2011})}\BibitemShut
  {NoStop}%
\bibitem [{\citenamefont {Tao}\ and\ \citenamefont {Thouless}(1983)}]{TT}%
  \BibitemOpen
  \bibfield  {author} {\bibinfo {author} {\bibfnamefont {R.}~\bibnamefont
  {Tao}}\ and\ \bibinfo {author} {\bibfnamefont {D.~J.}\ \bibnamefont
  {Thouless}},\ }\href {\doibase 10.1103/PhysRevB.28.1142} {\bibfield
  {journal} {\bibinfo  {journal} {Phys. Rev. B}\ }\textbf {\bibinfo {volume}
  {28}},\ \bibinfo {pages} {1142} (\bibinfo {year} {1983})}\BibitemShut
  {NoStop}%
\bibitem [{\citenamefont {Rezayi}\ and\ \citenamefont {Haldane}(1994)}]{RH}%
  \BibitemOpen
  \bibfield  {author} {\bibinfo {author} {\bibfnamefont {E.~H.}\ \bibnamefont
  {Rezayi}}\ and\ \bibinfo {author} {\bibfnamefont {F.~D.~M.}\ \bibnamefont
  {Haldane}},\ }\href {\doibase 10.1103/PhysRevB.50.17199} {\bibfield
  {journal} {\bibinfo  {journal} {Phys. Rev. B}\ }\textbf {\bibinfo {volume}
  {50}},\ \bibinfo {pages} {17199} (\bibinfo {year} {1994})}\BibitemShut
  {NoStop}%
\bibitem [{\citenamefont {Bergholtz}\ and\ \citenamefont
  {Karlhede}(2005)}]{BK1}%
  \BibitemOpen
  \bibfield  {author} {\bibinfo {author} {\bibfnamefont {E.~J.}\ \bibnamefont
  {Bergholtz}}\ and\ \bibinfo {author} {\bibfnamefont {A.}~\bibnamefont
  {Karlhede}},\ }\href {\doibase 10.1103/PhysRevLett.94.026802} {\bibfield
  {journal} {\bibinfo  {journal} {Phys. Rev. Lett.}\ }\textbf {\bibinfo
  {volume} {94}},\ \bibinfo {pages} {026802} (\bibinfo {year}
  {2005})}\BibitemShut {NoStop}%
\bibitem [{\citenamefont {Seidel}\ and\ \citenamefont {Lee}(2006)}]{SL}%
  \BibitemOpen
  \bibfield  {author} {\bibinfo {author} {\bibfnamefont {A.}~\bibnamefont
  {Seidel}}\ and\ \bibinfo {author} {\bibfnamefont {D.-H.}\ \bibnamefont
  {Lee}},\ }\href {\doibase 10.1103/PhysRevLett.97.056804} {\bibfield
  {journal} {\bibinfo  {journal} {Phys. Rev. Lett.}\ }\textbf {\bibinfo
  {volume} {97}},\ \bibinfo {pages} {056804} (\bibinfo {year}
  {2006})}\BibitemShut {NoStop}%
\bibitem [{\citenamefont {{Seidel}}\ and\ \citenamefont
  {{Lee}}(2007)}]{seidellee07}%
  \BibitemOpen
  \bibfield  {author} {\bibinfo {author} {\bibfnamefont {A.}~\bibnamefont
  {{Seidel}}}\ and\ \bibinfo {author} {\bibfnamefont {D.-H.}\ \bibnamefont
  {{Lee}}},\ }\href {\doibase 10.1103/PhysRevB.76.155101} {\bibfield  {journal}
  {\bibinfo  {journal} {\prb}\ }\textbf {\bibinfo {volume} {76}},\ \bibinfo
  {eid} {155101} (\bibinfo {year} {2007})},\ \Eprint
  {http://arxiv.org/abs/cond-mat/0611535} {cond-mat/0611535} \BibitemShut
  {NoStop}%
\bibitem [{\citenamefont {Bergholtz}\ and\ \citenamefont
  {Karlhede}(2006)}]{BK2}%
  \BibitemOpen
  \bibfield  {author} {\bibinfo {author} {\bibfnamefont {E.~J.}\ \bibnamefont
  {Bergholtz}}\ and\ \bibinfo {author} {\bibfnamefont {A.}~\bibnamefont
  {Karlhede}},\ }\href {http://stacks.iop.org/1742-5468/2006/i=04/a=L04001}
  {\bibfield  {journal} {\bibinfo  {journal} {Journal of Statistical Mechanics:
  Theory and Experiment}\ }\textbf {\bibinfo {volume} {2006}},\ \bibinfo
  {pages} {L04001} (\bibinfo {year} {2006})}\BibitemShut {NoStop}%
\bibitem [{\citenamefont {Bergholtz}\ and\ \citenamefont
  {Karlhede}(2008)}]{BK}%
  \BibitemOpen
  \bibfield  {author} {\bibinfo {author} {\bibfnamefont {E.~J.}\ \bibnamefont
  {Bergholtz}}\ and\ \bibinfo {author} {\bibfnamefont {A.}~\bibnamefont
  {Karlhede}},\ }\href {\doibase 10.1103/PhysRevB.77.155308} {\bibfield
  {journal} {\bibinfo  {journal} {Phys. Rev. B}\ }\textbf {\bibinfo {volume}
  {77}},\ \bibinfo {pages} {155308} (\bibinfo {year} {2008})}\BibitemShut
  {NoStop}%
\bibitem [{\citenamefont {Seidel}\ and\ \citenamefont {Yang}(2008)}]{SY08}%
  \BibitemOpen
  \bibfield  {author} {\bibinfo {author} {\bibfnamefont {A.}~\bibnamefont
  {Seidel}}\ and\ \bibinfo {author} {\bibfnamefont {K.}~\bibnamefont {Yang}},\
  }\href {\doibase 10.1103/PhysRevLett.101.036804} {\bibfield  {journal}
  {\bibinfo  {journal} {Phys. Rev. Lett.}\ }\textbf {\bibinfo {volume} {101}},\
  \bibinfo {pages} {036804} (\bibinfo {year} {2008})}\BibitemShut {NoStop}%
\bibitem [{\citenamefont {Ardonne}\ \emph {et~al.}(2008)\citenamefont
  {Ardonne}, \citenamefont {Bergholtz}, \citenamefont {Kailasvuori},\ and\
  \citenamefont {Wikberg}}]{ABKW}%
  \BibitemOpen
  \bibfield  {author} {\bibinfo {author} {\bibfnamefont {E.}~\bibnamefont
  {Ardonne}}, \bibinfo {author} {\bibfnamefont {E.~J.}\ \bibnamefont
  {Bergholtz}}, \bibinfo {author} {\bibfnamefont {J.}~\bibnamefont
  {Kailasvuori}}, \ and\ \bibinfo {author} {\bibfnamefont {E.}~\bibnamefont
  {Wikberg}},\ }\href {http://stacks.iop.org/1742-5468/2008/i=04/a=P04016}
  {\bibfield  {journal} {\bibinfo  {journal} {Journal of Statistical Mechanics:
  Theory and Experiment}\ }\textbf {\bibinfo {volume} {2008}},\ \bibinfo
  {pages} {P04016} (\bibinfo {year} {2008})}\BibitemShut {NoStop}%
\bibitem [{\citenamefont {Bergholtz}\ \emph {et~al.}(2008)\citenamefont
  {Bergholtz}, \citenamefont {Hansson}, \citenamefont {Hermanns}, \citenamefont
  {Karlhede},\ and\ \citenamefont {Viefers}}]{BHHKV}%
  \BibitemOpen
  \bibfield  {author} {\bibinfo {author} {\bibfnamefont {E.~J.}\ \bibnamefont
  {Bergholtz}}, \bibinfo {author} {\bibfnamefont {T.~H.}\ \bibnamefont
  {Hansson}}, \bibinfo {author} {\bibfnamefont {M.}~\bibnamefont {Hermanns}},
  \bibinfo {author} {\bibfnamefont {A.}~\bibnamefont {Karlhede}}, \ and\
  \bibinfo {author} {\bibfnamefont {S.}~\bibnamefont {Viefers}},\ }\href
  {\doibase 10.1103/PhysRevB.77.165325} {\bibfield  {journal} {\bibinfo
  {journal} {Phys. Rev. B}\ }\textbf {\bibinfo {volume} {77}},\ \bibinfo
  {pages} {165325} (\bibinfo {year} {2008})}\BibitemShut {NoStop}%
\bibitem [{\citenamefont {Seidel}(2010)}]{seidel10}%
  \BibitemOpen
  \bibfield  {author} {\bibinfo {author} {\bibfnamefont {A.}~\bibnamefont
  {Seidel}},\ }\href {\doibase 10.1103/PhysRevLett.105.026802} {\bibfield
  {journal} {\bibinfo  {journal} {Phys. Rev. Lett.}\ }\textbf {\bibinfo
  {volume} {105}},\ \bibinfo {pages} {026802} (\bibinfo {year}
  {2010})}\BibitemShut {NoStop}%
\bibitem [{\citenamefont {L\"auchli}\ \emph {et~al.}(2010)\citenamefont
  {L\"auchli}, \citenamefont {Bergholtz}, \citenamefont {Suorsa},\ and\
  \citenamefont {Haque}}]{LBSH}%
  \BibitemOpen
  \bibfield  {author} {\bibinfo {author} {\bibfnamefont {A.~M.}\ \bibnamefont
  {L\"auchli}}, \bibinfo {author} {\bibfnamefont {E.~J.}\ \bibnamefont
  {Bergholtz}}, \bibinfo {author} {\bibfnamefont {J.}~\bibnamefont {Suorsa}}, \
  and\ \bibinfo {author} {\bibfnamefont {M.}~\bibnamefont {Haque}},\ }\href
  {\doibase 10.1103/PhysRevLett.104.156404} {\bibfield  {journal} {\bibinfo
  {journal} {Phys. Rev. Lett.}\ }\textbf {\bibinfo {volume} {104}},\ \bibinfo
  {pages} {156404} (\bibinfo {year} {2010})}\BibitemShut {NoStop}%
\bibitem [{\citenamefont {Seidel}\ and\ \citenamefont {Yang}(2011)}]{SY11}%
  \BibitemOpen
  \bibfield  {author} {\bibinfo {author} {\bibfnamefont {A.}~\bibnamefont
  {Seidel}}\ and\ \bibinfo {author} {\bibfnamefont {K.}~\bibnamefont {Yang}},\
  }\href {\doibase 10.1103/PhysRevB.84.085122} {\bibfield  {journal} {\bibinfo
  {journal} {Phys. Rev. B}\ }\textbf {\bibinfo {volume} {84}},\ \bibinfo
  {pages} {085122} (\bibinfo {year} {2011})}\BibitemShut {NoStop}%
\bibitem [{\citenamefont {Papi\ifmmode~\acute{c}\else
  \'{c}\fi{}}(2014)}]{papic}%
  \BibitemOpen
  \bibfield  {author} {\bibinfo {author} {\bibfnamefont {Z.}~\bibnamefont
  {Papi\ifmmode~\acute{c}\else \'{c}\fi{}}},\ }\href {\doibase
  10.1103/PhysRevB.90.075304} {\bibfield  {journal} {\bibinfo  {journal} {Phys.
  Rev. B}\ }\textbf {\bibinfo {volume} {90}},\ \bibinfo {pages} {075304}
  (\bibinfo {year} {2014})}\BibitemShut {NoStop}%
\bibitem [{\citenamefont {Weerasinghe}\ and\ \citenamefont
  {Seidel}(2014)}]{WS}%
  \BibitemOpen
  \bibfield  {author} {\bibinfo {author} {\bibfnamefont {A.}~\bibnamefont
  {Weerasinghe}}\ and\ \bibinfo {author} {\bibfnamefont {A.}~\bibnamefont
  {Seidel}},\ }\href {\doibase 10.1103/PhysRevB.90.125146} {\bibfield
  {journal} {\bibinfo  {journal} {Phys. Rev. B}\ }\textbf {\bibinfo {volume}
  {90}},\ \bibinfo {pages} {125146} (\bibinfo {year} {2014})}\BibitemShut
  {NoStop}%
\bibitem [{\citenamefont {{Bernevig}}\ and\ \citenamefont
  {{Regnault}}(2009)}]{regnault}%
  \BibitemOpen
  \bibfield  {author} {\bibinfo {author} {\bibfnamefont {B.~A.}\ \bibnamefont
  {{Bernevig}}}\ and\ \bibinfo {author} {\bibfnamefont {N.}~\bibnamefont
  {{Regnault}}},\ }\href {\doibase 10.1103/PhysRevLett.103.206801} {\bibfield
  {journal} {\bibinfo  {journal} {Physical Review Letters}\ }\textbf {\bibinfo
  {volume} {103}},\ \bibinfo {eid} {206801} (\bibinfo {year} {2009})},\ \Eprint
  {http://arxiv.org/abs/0902.4320} {arXiv:0902.4320} \BibitemShut {NoStop}%
\bibitem [{Note1()}]{Note1}%
  \BibitemOpen
  \bibinfo {note} {This corresponds to the familiar multiplication with
  symmetric polynomials in first quantized language.}\BibitemShut {Stop}%
\bibitem [{\citenamefont {Nakamura}\ \emph {et~al.}(2012)\citenamefont
  {Nakamura}, \citenamefont {Wang},\ and\ \citenamefont
  {Bergholtz}}]{Nakamura}%
  \BibitemOpen
  \bibfield  {author} {\bibinfo {author} {\bibfnamefont {M.}~\bibnamefont
  {Nakamura}}, \bibinfo {author} {\bibfnamefont {Z.-Y.}\ \bibnamefont {Wang}},
  \ and\ \bibinfo {author} {\bibfnamefont {E.~J.}\ \bibnamefont {Bergholtz}},\
  }\href {\doibase 10.1103/PhysRevLett.109.016401} {\bibfield  {journal}
  {\bibinfo  {journal} {Phys. Rev. Lett.}\ }\textbf {\bibinfo {volume} {109}},\
  \bibinfo {pages} {016401} (\bibinfo {year} {2012})}\BibitemShut {NoStop}%
\bibitem [{Note2()}]{Note2}%
  \BibitemOpen
  \bibinfo {note} {See. Ref. \protect \rev@citealpnum {ortiz} for a
  second-quantized proof.}\BibitemShut {Stop}%
\bibitem [{\citenamefont {Read}(1989)}]{readOP}%
  \BibitemOpen
  \bibfield  {author} {\bibinfo {author} {\bibfnamefont {N.}~\bibnamefont
  {Read}},\ }\href {\doibase 10.1103/PhysRevLett.62.86} {\bibfield  {journal}
  {\bibinfo  {journal} {Phys. Rev. Lett.}\ }\textbf {\bibinfo {volume} {62}},\
  \bibinfo {pages} {86} (\bibinfo {year} {1989})}\BibitemShut {NoStop}%
\bibitem [{\citenamefont {Mazaheri}\ \emph {et~al.}(2015)\citenamefont
  {Mazaheri}, \citenamefont {Ortiz}, \citenamefont {Nussinov},\ and\
  \citenamefont {Seidel}}]{mazaheri_et_al}%
  \BibitemOpen
  \bibfield  {author} {\bibinfo {author} {\bibfnamefont {T.}~\bibnamefont
  {Mazaheri}}, \bibinfo {author} {\bibfnamefont {G.}~\bibnamefont {Ortiz}},
  \bibinfo {author} {\bibfnamefont {Z.}~\bibnamefont {Nussinov}}, \ and\
  \bibinfo {author} {\bibfnamefont {A.}~\bibnamefont {Seidel}},\ }\href
  {\doibase 10.1103/PhysRevB.91.085115} {\bibfield  {journal} {\bibinfo
  {journal} {Phys. Rev. B}\ }\textbf {\bibinfo {volume} {91}},\ \bibinfo
  {pages} {085115} (\bibinfo {year} {2015})}\BibitemShut {NoStop}%
\bibitem [{Note3()}]{Note3}%
  \BibitemOpen
  \bibinfo {note} {Let $[j]_0=1$, $[j]_m=j(j-1)\protect \dotsc (j-m+1)$ for
  $m>0$. Then $\DOTSB \sum@ \slimits@ _{{0\leq j\leq M}} [j]_m (-1)^j {M
  \atopwithdelims ()j}=\DOTSB \sum@ \slimits@ _{{m\leq j\leq M}} \protect \frac
  {j!}{(j-m)!} (-1)^j {M \atopwithdelims ()j} =(-1)^m [M]_m (1-1)^{M-m}
  \protect \tmspace +\thinmuskip {.1667em}.$ This gives $0$ for $0\leq m < M$.
  Since $[j]_m$ is clearly an $m$th degree polynomial in $j$, we can make new
  linear combinations of the latter identities to obtain Eq.~(\ref {jm}).
  Extending consideration to $m=M$ gives all the ingredients for the
  interesting identity \cite {Ruiz96} $\DOTSB \sum@ \slimits@ _{j=0}^M
  (-1)^j{M\atopwithdelims ()j}(x-j)^M= M! \hskip 1em\relax \forall
  x$.}\BibitemShut {Stop}%
\bibitem [{\citenamefont {Ruiz}(1996)}]{Ruiz96}%
  \BibitemOpen
  \bibfield  {author} {\bibinfo {author} {\bibfnamefont {S.~M.}\ \bibnamefont
  {Ruiz}},\ }\href {http://www.jstor.org/stable/3618534} {\bibfield  {journal}
  {\bibinfo  {journal} {The Mathematical Gazette}\ }\textbf {\bibinfo {volume}
  {80}},\ \bibinfo {pages} {pp. 579} (\bibinfo {year} {1996})}\BibitemShut
  {NoStop}%
\bibitem [{\citenamefont {Cirac}\ and\ \citenamefont {Sierra}(2010)}]{cirac}%
  \BibitemOpen
  \bibfield  {author} {\bibinfo {author} {\bibfnamefont {J.~I.}\ \bibnamefont
  {Cirac}}\ and\ \bibinfo {author} {\bibfnamefont {G.}~\bibnamefont {Sierra}},\
  }\href {\doibase 10.1103/PhysRevB.81.104431} {\bibfield  {journal} {\bibinfo
  {journal} {Phys. Rev. B}\ }\textbf {\bibinfo {volume} {81}},\ \bibinfo
  {pages} {104431} (\bibinfo {year} {2010})}\BibitemShut {NoStop}%
\end{thebibliography}%

\end{document}